\documentclass[english,aps,prb,twocolumn,superscriptaddress,showpacs,showkeys]{revtex4-1}
\usepackage[T1]{fontenc}
\usepackage{xcolor}
\usepackage{pdfcolmk}
\usepackage{babel}
\usepackage{array}
\usepackage{booktabs}
\usepackage{units}
\usepackage{textcomp}
\usepackage{multirow}
\usepackage{amstext}
\usepackage{graphicx}
\PassOptionsToPackage{normalem}{ulem}
\usepackage{ulem}
\usepackage[unicode=true,
 bookmarks=true,bookmarksnumbered=false,bookmarksopen=false,
 breaklinks=false,pdfborder={0 0 1},backref=false,colorlinks=false]
 {hyperref}
\hypersetup{pdftitle={Evolution of ferromagnetism and nFL state with doping: the case of Ru doped UCoGe},
 pdfauthor={M. Valiska, J. Pospisil, V. Sechovsky, M. Divis},
 pdfkeywords={Keywords: UCoGe, URuGe, Quantum critical point, non-Fermi liquid behavior, ferromagnetism}}
\usepackage{breakurl}

\makeatletter

\@ifundefined{textcolor}{}
{%
 \definecolor{BLACK}{gray}{0}
 \definecolor{WHITE}{gray}{1}
 \definecolor{RED}{rgb}{1,0,0}
 \definecolor{GREEN}{rgb}{0,1,0}
 \definecolor{BLUE}{rgb}{0,0,1}
 \definecolor{CYAN}{cmyk}{1,0,0,0}
 \definecolor{MAGENTA}{cmyk}{0,1,0,0}
 \definecolor{YELLOW}{cmyk}{0,0,1,0}
}

\makeatother

\begin{document}

\title{Evolution of ferromagnetic and non-Fermi liquid state with doping:
the case of Ru doped UCoGe }

\author{Michal Vali\v{s}ka}

\email{michal.valiska@gmail.com}

\selectlanguage{english}%

\affiliation{Faculty of Mathematics and Physics, Charles University, DCMP, Ke
Karlovu 5, CZ-12116 Praha 2, Czech Republic}

\author{Ji\v{r}í Pospí\v{s}il}

\affiliation{Advanced Science Research Center, Japan Atomic Energy Agency, Tokai,
Ibaraki, 319-1195, Japan}

\author{Martin Divi\v{s}}

\author{Jan Prokle\v{s}ka}

\author{Vladimír Sechovský}

\affiliation{Faculty of Mathematics and Physics, Charles University, DCMP, Ke
Karlovu 5, CZ-12116 Praha 2, Czech Republic}

\author{Mohsen M. Abd-Elmeguid}

\affiliation{II. Physikalisches Institut, Universität zu Köln, 0937 Köln, Germany}
\begin{abstract}
We have investigated the impact of Ru substitution for Co on the behavior
of the ferromagnetic superconductor UCoGe by performing x-ray diffraction,
magnetization, specific heat and electrical resistivity measurements
on polycrystalline samples of the $\mathrm{UCo}_{1-x}\mathrm{Ru}_{x}\mathrm{Ge}$
series ($0\geq x\leq0.9$). The initial Ru substitution up to $x\approx0.1$
leads to a simultaneous sharp increase of the Curie temperature and
spontaneous magnetization up to maximum values of $T_{\mathrm{C}}=\unit[8.6]{K}$
and $M_{\mathrm{S}}=\unit[0.1]{\mu_{\mathrm{B}}}$ per formula unit,
respectively, whereas superconductivity vanishes already for $x\approx0.03$.
Further increase of the Ru content beyond $x\approx0.1$ leads to
a precipitous decrease of both, $T_{\mathrm{C}}$ and $M_{\mathrm{S}}$
towards a ferromagnetic quantum critical point (QCP) at $x_{\mathrm{cr}}=0.31$.
Consequently the $T-x$ magnetic phase diagram consists of a well-developed
ferromagnetic dome. We discuss the evolution of ferromagnetism with
$x$ on the basis of band structure changes due to varying 5$f$-ligand
hybridization. This scenario is supported by the results of electronic
structure calculations and consideration of the simplified periodic
Anderson model. The analysis of the temperature dependencies of the
electrical resistivity and heat capacity at low temperatures of the
samples in the vicinity of the QCP reveals a non-Fermi liquid behavior
and assigns the ferromagnetic quantum phase transition to be most
likely of a continuous Hertz-Millis type.
\end{abstract}



\maketitle

\section{Introduction}

The phenomena emerging near a quantum critical point (QCP) belong
to the most intensively studied topics of condensed matter physics.
Diligent research in this field continuously brings brand new materials
carrying completely novel properties. Such progress boosts development
of new theoretical approaches describing electron correlations in
these systems. A specific group of those intriguing materials comprises
the uranium based ferromagnetic superconductors (FM SC) $\mathrm{UGe_{2}}$\cite{Saxena2000,Huxley2001},
URhGe\cite{Aoki2001} and UCoGe\cite{Huy2007}. In these compounds
superconductivity and itinerant ferromagnetism are carried by the
same uranium 5$f$ electrons. It is a novelty distinguishing them
from previously reported $\mathrm{ZrZn_{2}}$\cite{Uhlarz2004}. $\mathrm{UGe_{2}}$,
the first discovered case, is a model example of superconductivity
(SC) induced by external pressure. Here SC appears and reaches a maximum
$T_{\mathrm{SC}}$ on a boundary between two different FM phases under
high pressure. URhGe and UCoGe are ambient pressure FM SC where both
phenomena naturally coexist. A lot of effort both in theory and experiment
has been done to explore the underlying mechanisms of the coexistence
of FM and SC. Ferromagnetic spin fluctuations which appear in the
vicinity of the QCP have been considered as the main essence for inducing
unconventional spin-triplet SC state\cite{deNijs2008,Stock2011,Hatorri2012}.

Quantum phase transitions (QPTs) were experimentally studied for a
broad spectrum of materials like high-$T_{\mathrm{C}}$ superconductors\cite{Marel2003},
ordinary metals\cite{Pfleiderer1997} or heavy-fermion compounds\cite{Schroder2000}.
Most of such investigations have been carried out on antiferromagnets
which by rule exhibit second-order QCP. Prominent examples are $\mathrm{CeCu}_{6-x}\mathrm{Au}_{x}$
with an antiferromagnetic quantum critical point (AF QCP) which is
induced by chemical doping\cite{Lohneysen1994} or $\mathrm{YbRh_{2}Si_{2}}$
where the AF QCP is achieved by applying external magnetic field\cite{Custers2003}.
Studies of quantum criticality in ferromagnets have been less frequent
and manifest that here the situation may be much more complex. The
ferromagnetic phase transition at finite Curie temperature ($T_{\mathrm{C}}$)
is by rule of a second order type. $T_{\mathrm{C}}$ of itinerant
electron ferromagnets is often easily suppressed to $\unit[0]{K}$
by external pressure $p$ or chemical composition $x$. However, detailed
experimental investigation of archetypal ferromagnetic metals such
as MnSi\cite{Pfleiderer1997}, $\mathrm{ZrZn_{2}}$\cite{Uhlarz2004}
or $\mathrm{UGe_{2}}$\cite{Huxley2001,Pfleiderer2002} has revealed
that the ferromagnetic phase is suppressed to zero temperature at
a first-order transition which would mean that no QCP is observed.
This can be elucidated theoretically either in terms of additional
fermionic modes which may couple to the critical ferromagnetic fluctuations
driving the phase transition to a first-order type\cite{Belitz1999}
or that a first-order magnetic phase transition may be induced by
strong magneto-elastic coupling\cite{Mineev2011}. No generic scenario
can be drawn for the QPTs of the above mentioned materials because
of rather individually different phenomena appearing in the quantum
critical region. In particular MnSi becomes long-period helimagnet
(showing ferromagnetism only locally) in which the thermal phase transition
is weakly first-order\cite{Janoschek2013}, $\mathrm{UGe_{2}}$ exhibits
strong uniaxial anisotropy\cite{Menovsky1983} and $\mathrm{ZrZn_{2}}$
exhibits a marginal Fermi liquid ground state\cite{Sutherland2012}.

UCoGe, the subject of the present study, is unique in the group of
FM SC due to the much lower energy scale on which the magnetism appears\cite{Aoki2012}.
The low Curie temperature of UCoGe is only $\unit[3]{K}$\cite{Huy2007,Gasparini2010a}
together with the tiny spontaneous magnetization of $\unit[0.03]{\mu_{\mathrm{B}}}$
per formula unit (f.u.) indicate that UCoGe is close to a ferromagnetic
instability\cite{Aoki2009}. It has been observed, however, that the
Ru and Fe substitution for Co rapidly stabilizes the ferromagnetic
state\cite{Pospisil2009}, despite the fact that URuGe and UFeGe behave
like Pauli paramagnets down to the lowest temperatures\cite{Troc1988}.
Similar increase of $T_{\mathrm{C}}$ was reported in the case of
the initial substitution of Co and Ru for Rh in URhGe\cite{Huy2009,Huy2007b}
with the development of a non-Fermi liquid (NFL) state on the higher
doping boundary of the FM dome\cite{Huy2007b}. These observations
motivated us to inspect the development of the magnetic as well as
electrical and thermal transport properties in the $\mathrm{UCo}_{1-x}\mathrm{Ru}_{x}\mathrm{Ge}$
series over the entire concentration range ($0\geq x\leq0.9$). Our
study is based on extensive investigation of the crystal structure,
magnetization, AC magnetic susceptibility, specific heat and electrical
resistance of numerous polycrystalline samples with various Ru content.
The results are discussed and compared with theoretical calculations
and related models considering the leading role of the 5$f$-ligand
hybridization.

\section{Experimental Details}

In order to study the development of the magnetic state in the $\mathrm{UCo}_{1-x}\mathrm{Ru}_{x}\mathrm{Ge}$
system we have prepared a series of polycrystalline samples with different
Ru concentrations $x$ between 0 and 0.9. All samples were prepared
by arc-melting of the stoichiometric amounts of the elements (purity
of Co 4N5, Ge 6N and Ru 3N5). U was purified by the Solid State Electrotransport
technique (SSE)\cite{Pospisil2011} following previous experience
with preparation of UCoGe\cite{Pospisil2011}. The arc melting process
was realized under protective Ar (6N purity) atmosphere on a water
cooled Cu crucible. Each sample was three times turned upside down
and subsequently re-melted in order to achieve the best homogeneity.
All samples were separately wrapped into a Ta foil (99.99\%), sealed
in a quartz tube under the vacuum of $\unit[1\cdot10^{-6}]{mbar}$
, subsequently annealed at $\unit[885]{\text{\textdegree C}}$ for
14 days and then slowly cooled down to room temperature to avoid creation
of the internal stresses. Each sample was characterized by X-ray powder
diffraction (XRPD) at room temperature on a Bruker D8 Advance diffractometer.
The obtained data were evaluated by Rietveld technique\cite{Rietveld1969}
using FullProf/WinPlotr software\cite{Rodriguez-Carvajal1993,Roisnel2000}
with respect to the previously published crystallographic data of
the UCoGe\cite{Canepa1996} and URuGe\cite{Troc1988} compound. The
chemical composition of our samples was verified by a scanning electron
microscope (SEM) Tescan Mira I LMH equipped with an energy dispersive
X-ray detector (EDX) Bruker AXS. Samples were afterward properly shaped
for individual measurements with a fine wire saw to prevent induction
of additional stresses and lattice defects. The electrical resistivity
($\rho$) was measured by the 4-probe method on bar-shape samples
($\unit[1\times0.5\times4]{mm^{3}}$) and heat-capacity ($C_{\mathrm{p}}$)
measurements were performed on thin plates ($\unit[2\times2\times0.2]{mm^{3}}$)
by the relaxation method on PPMS9T and PPMS14T devices using a $^{3}\mathrm{He}$
insert. Magnetization ($M$) measurements were done on cubic samples
($\unit[2\times2\times2]{mm^{3}}$) using a MPMS7T device. The magnetization
was evaluated in $\mu_{\mathrm{B}}/\mathrm{f.u.}$. For simplicity
we omit \textquotedblleft $/\mathrm{f.u.}$\textquotedblright{} everywhere
throughout the paper.

The electronic structure calculations were performed on the basis
of the density-functional theory (DFT) within the local-spin-density
approximation (LSDA)\cite{Perdew1992} and the generalized gradient
approximation (GGA)\cite{Perdew1996}. For this calculations we have
used the full-potential augmented-plane-wave together with the local-orbitals
method (APW+lo) as a part of the latest version (WIEN2k) of the original
WIEN code\cite{schwarz2002}.

\section{Results}

\subsection{X-ray diffraction }

Both UCoGe and URuGe crystalize in the orthorhombic TiNiSi-type structure
(space group Pnma)\cite{Canepa1996,Troc1988} with the room-temperature
cell parameters $a=\unit[6.852]{\textrm{\AA}}$, $b=\unit[4.208]{\textrm{\AA}}$,
$c=\unit[7.226]{\textrm{\AA}}$ and $a=\unit[6.678]{\textrm{\AA}}$,
$b=\unit[4.359]{\textrm{\AA}}$, $c=\unit[7.539]{\textrm{\AA}}$,
respectively\cite{Canepa1996,Troc1988}. The unit cell volume of UCoGe
($V=\unit[208.3]{\textrm{\AA}^{3}}$)\cite{Canepa1996} is about 5\%
smaller than that of the URuGe compound ($V=\unit[219.5]{\textrm{\AA}^{3}}$)\cite{Troc1988}.
The XRPD patterns confirmed the orthorhombic TiNiSi-type structure
of samples over the entire concentration range of the $\mathrm{UCo}_{1-x}\mathrm{Ru}_{x}\mathrm{Ge}$
series.

\begin{figure}
\begin{centering}
\includegraphics[scale=0.5]{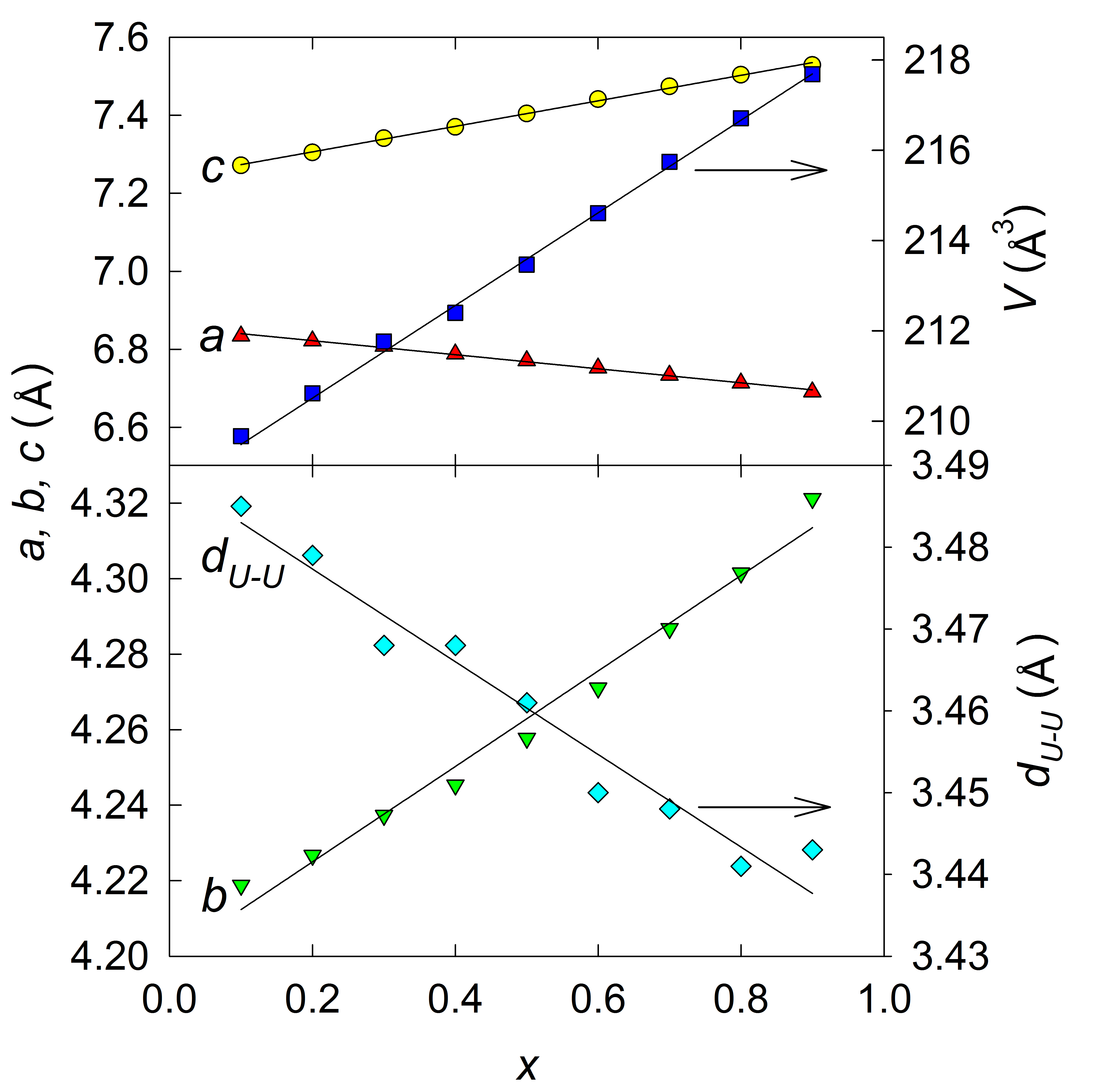}
\par\end{centering}

\protect\caption{\label{fig:(Color-online)--}(Color online) - Concentration dependence
of the lattice parameters and the unit cell volume of the $\mathrm{UCo}_{1-x}\mathrm{Ru}_{x}\mathrm{Ge}$
samples. The lines serve as guides to the eye.}

\end{figure}

The evaluated lattice parameters are listed in Table \ref{tab:Unit-cell-parameters}.
The concentration dependence of all three lattice parameters and the
unit volume reveals a linear behavior, i.e. obeying Vegard\textquoteright s
law\cite{Vegard1921} (see Fig. \ref{fig:(Color-online)--}).

\begin{table}
\protect\caption{\label{tab:Unit-cell-parameters}The lattice parameters and the unit
cell volume of the $\mathrm{UCo}_{1-x}\mathrm{Ru}_{x}\mathrm{Ge}$
samples as obtained from the refinement of the X-ray powder diffraction
patterns.}

\begin{ruledtabular}

\begin{centering}
\begin{tabular}{ccccc}
$x$ & $\unit[a]{\left(\textrm{\AA}\right)}$ & $\unit[b]{\left(\textrm{\AA}\right)}$ & $\unit[c]{\left(\textrm{\AA}\right)}$ & $\unit[V]{\left(\textrm{\AA}^{3}\right)}$\tabularnewline
\midrule
0.10 & 6.8344 & 4.2188 & 7.2717 & 209.6671\tabularnewline
0.20 & 6.8216 & 4.2267 & 7.3048 & 210.6173\tabularnewline
0.21 & 6.8204 & 4.2261 & 7.3026 & 210.4879\tabularnewline
0.22 & 6.8189 & 4.2279 & 7.3079 & 210.6840\tabularnewline
0.23 & 6.8178 & 4.2272 & 7.3085 & 210.6325\tabularnewline
0.24 & 6.8131 & 4.2280 & 7.3145 & 210.6999\tabularnewline
0.25 & 6.8150 & 4.2320 & 7.3229 & 211.2003\tabularnewline
0.26 & 6.8269 & 4.2385 & 7.3347 & 212.2355\tabularnewline
0.27 & 6.8203 & 4.2363 & 7.3314 & 211.8247\tabularnewline
0.28 & 6.8205 & 4.2390 & 7.3392 & 212.1890\tabularnewline
0.29 & 6.8133 & 4.2373 & 7.3389 & 211.8711\tabularnewline
0.30 & 6.8077 & 4.2373 & 7.3413 & 211.7662\tabularnewline
0.40 & 6.7880 & 4.2454 & 7.3704 & 212.3984\tabularnewline
0.50 & 6.7709 & 4.2577 & 7.4046 & 213.4669\tabularnewline
0.60 & 6.7522 & 4.2710 & 7.4416 & 214.6050\tabularnewline
0.70 & 6.7336 & 4.2868 & 7.4741 & 215.7451\tabularnewline
0.80 & 6.7137 & 4.3015 & 7.5041 & 216.7105\tabularnewline
0.90 & 6.6909 & 4.3212 & 7.5290 & 217.6849\tabularnewline
\end{tabular}
\par\end{centering}

\end{ruledtabular}
\end{table}

While the lattice parameters $b$ and $c$ increase with increasing
$x$, the lattice parameter $a$ simultaneously decreases. The volume
expansion seems to reflect the increase of the covalent radii from
Co (126 pm) to Ru (146 pm)\cite{Cordero2008}. Refinement of the diffraction
patterns showed, that the Ru atoms really substitute the Co ones on
their sites.

Although the unit cell volume expands with increasing Ru concentration
the distance between the nearest-neighbor U ions $d_{\mathrm{U-U}}$
contracts (see Fig.\ref{fig:(Color-online)--}). This result is not
surprising because the $d_{\mathrm{U-U}}$ lines form a chain meandering
along the $a$-axis.

\subsection{Magnetization and AC-Susceptibility}

We have measured the magnetization of each sample as a function of
temperature and applied magnetic field. The values of $M_{\mathrm{S}}$
have been estimated from the magnetization curves measured at $\unit[1.85]{K}$
(the lowest available temperature in our MPMS7T) by extrapolating
the magnetization from high magnetic fields to $\unit[0]{T}$.

\begin{figure}
\begin{centering}
\includegraphics[scale=0.5]{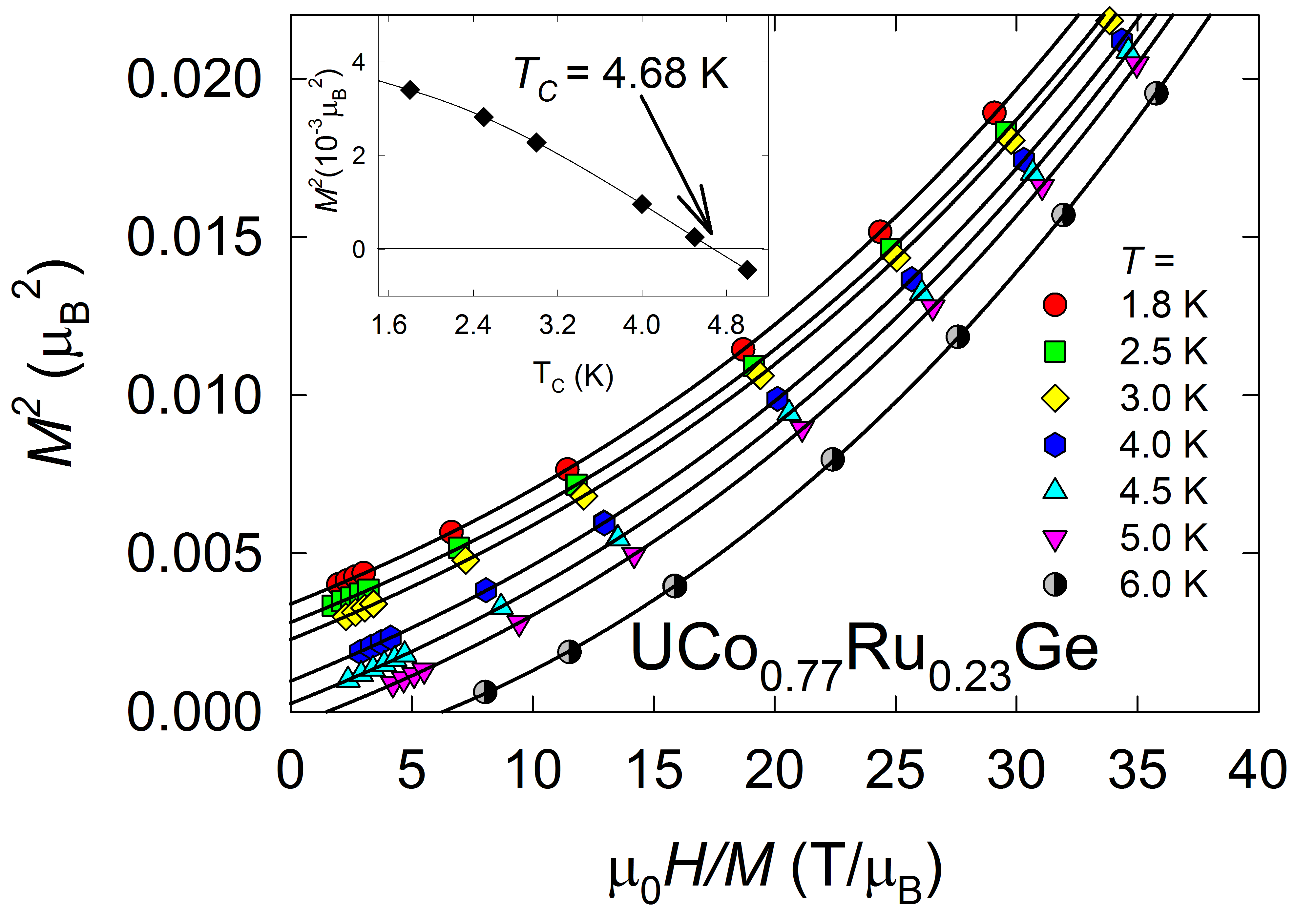}
\par\end{centering}

\protect\caption{\label{fig:(Color-online)-Arrott}(Color online) - Arrott plots for
the $\mathrm{UCo_{0.77}Ru_{0.23}Ge}$ compound. Solid lines are the
third order polynomial functions. The inset shows that $T_{\mathrm{C}}$
is taken as the value for which the intersection with the $M^{2}$
axis would be zero.}
\end{figure}

The values of $T_{\mathrm{C}}$ have been determined by several methods.
Arrott plot analysis of magnetization data is widely considered as
the most reliable method\cite{Arrott1957}. For this purpose the magnetization
curves were measured at several temperatures in the vicinity of the
expected $T_{\mathrm{C}}$. The Arrott plots obtained from our magnetization
data are strongly nonlinear. These curves can be approximated by a
third degree polynomial function (see a model example in Fig. \ref{fig:(Color-online)-Arrott}).
$T_{\mathrm{C}}$ is determined as the temperature of the Arrott plot
isotherm that would cross the $M^{2}$ axis at 0. An example of the
relevant construction is shown in the inset of Fig. \ref{fig:(Color-online)-Arrott}.

\begin{figure}
\begin{centering}
\includegraphics[scale=0.5]{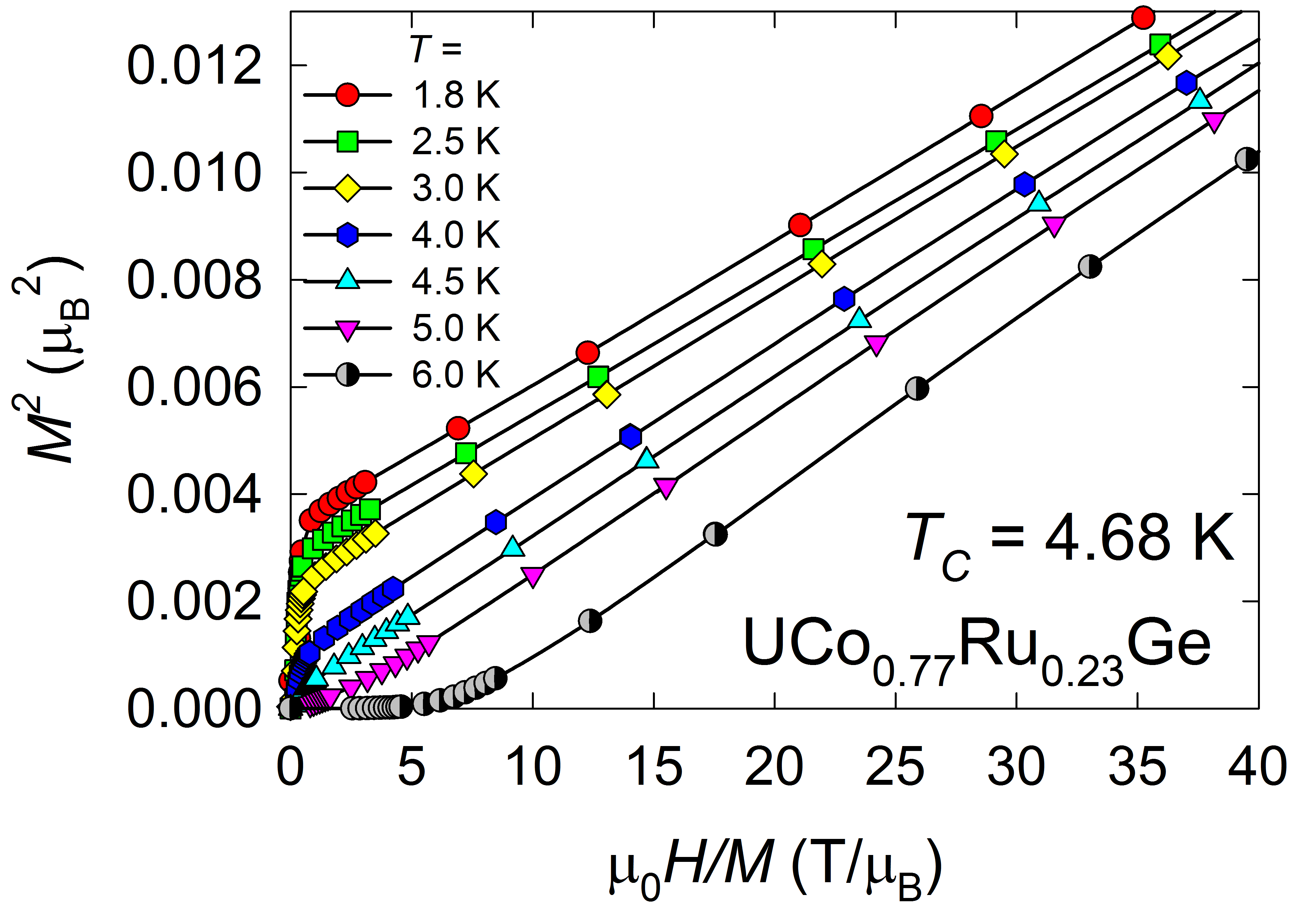}
\par\end{centering}

\protect\caption{\label{fig:(Color-online)---2}(Color online) - Revised Arrott plots
after subtraction of the linear term with the slope $a=\unit[0.006]{\mu_{\mathrm{B}}\cdot T^{-1}}$
from the magnetization data measured on the $\mathrm{UCo_{0.77}Ru_{0.23}Ge}$
sample.}
\end{figure}

The nonlinearity of the Arrott plots (the cubic $M^{2}$ vs $H/M$
dependence) suggests presence of a magnetization component linearly
dependent on the magnetic field. This is related to the fact that
UCoGe and the other U$TX$ compounds crystallizing in the orthorhombic
TiNiSi-type structure exhibit strong uniaxial anisotropy with easy
magnetization direction along the $c$-axis. The hard magnetization
directions within the $a-b$ plane are characteristic by a weak temperature-independent
paramagnetic response with the magnetization proportional to the magnetic
field. We have observed the same type of magnetocrystalline anisotropy
for the ferromagnetic $\mathrm{UCo}_{1-x}\mathrm{Ru}_{x}\mathrm{Ge}$
single crystals which we have grown as a part of another study (see
Ref.\cite{Valiska2015}). Consequently the polycrystalline samples
should show a corresponding linear component also in the ferromagnetic
state. By subtracting a suitable linear term from measured magnetization
data we obtain the corrected magnetization values $M^{*}=M-a\cdot\mu_{0}\cdot H$.
For $a=\unit[0.006]{\mu_{\mathrm{B}}\cdot T^{-1}}$ the Arrott plots
$M^{2}$ vs $H/M^{*}$ are indeed linear except the low-field part
due to low-field magnetization processes and influence of a demagnetization
field as can be seen for example in the case of the $\mathrm{UCo_{0.77}Ru_{0.23}Ge}$
sample in Fig. \ref{fig:(Color-online)---2}. 

The obtained $T_{\mathrm{C}}$ and $M_{\mathrm{S}}$ values are listed
for all samples in Table \ref{tab:Values-of-the} and plotted in the
complex phase diagram in Fig. \ref{fig:(Color-online)-Panel}(a).
$T_{\mathrm{C}}$ steeply increases with the initial Ru substitutions
for Co which is in agreement with the results published in previous
work\cite{Pospisil2009}. This trend terminates at $x_{\mathrm{max}}\approx0.1$
where the ordering temperature reaches a maximum value of $T_{\mathrm{C,max}}\approx\unit[8.6]{K}$.
This value is almost three times higher than $T_{\mathrm{C}}=\unit[3]{K}$
of the parent compound\cite{Huy2007} and is comparable with the value
found by Huang et al. in the case of the corresponding substitution
of Fe for Co in UCoGe\cite{Huang2013}. Increasing Ru concentration
beyond $x\approx0.1$ is accompanied by a simultaneous decrease of
$T_{\mathrm{C}}$ and $M_{\mathrm{S}}$ towards zero at the critical
concentration $x_{\mathrm{cr}}\approx0.31$. Thus, the ferromagnetic
dome of the concentration dependence of $T_{\mathrm{C}}$ in the $T-x$
magnetic phase diagram is intimately connected with a corresponding
change of $M_{\mathrm{S}}$ (see Fig. \ref{fig:(Color-online)-Panel}(a)).

\begin{table}
\protect\caption{\label{tab:Values-of-the}Values of the spontaneous magnetization
$M_{\mathrm{S}}$ and Curie temperature derived from Arrott plot analysis
($T_{\mathrm{C,Arrott}}$), temperature dependence of AC susceptibility
($T_{\mathrm{C,\chi}}$), magnetization ($T_{\mathrm{C},M}$) and
specific heat ($T_{\mathrm{C},C_{\mathrm{p}}}$), and Sommerfeld coefficient
($\gamma$) as determined from the specific heat data at low temperatures
for samples with various concentration of Ru ($x$).}

\begin{ruledtabular}

\begin{centering}
\begin{tabular}{ccccccc}
\multirow{2}{*}{$x$} & $M_{\mathrm{S}}$  & $T_{\mathrm{C,Arrott}}$ & $T_{\mathrm{C,\chi}}$ & $T_{\mathrm{C},M}$ & $T_{\mathrm{C},C_{\mathrm{p}}}$ & $\gamma$\tabularnewline
 & $\left(\mu_{\mathrm{B}}\right)$ & $\left(K\right)$ & $\left(K\right)$ & $\left(K\right)$ & $\left(K\right)$ & $\left(\frac{\mathrm{mJ}}{\mathrm{mol\cdot K^{2}}}\right)$\tabularnewline
\midrule
0 & 0.0300 & - & - & 2.50 & - & -\tabularnewline
0.01 & 0.0330 & 4.20 & - & 4.00 & - & -\tabularnewline
0.05 & 0.0750 & 8.30 & - & 7.50 & - & -\tabularnewline
0.10 & 0.1060 & 8.62 & - & 8.20 & 8.60 & 0.0861\tabularnewline
0.20 & 0.0540 & 5.70 & - & 5.40 & 5.70 & 0.1066\tabularnewline
0.21 & 0.0580 & 5.70 & 5.20 & 5.40 & 5.90 & 0.1100\tabularnewline
0.22 & 0.0594 & 5.01 & 4.70 & 5.00 & 5.30 & 0.1133\tabularnewline
0.23 & 0.0568 & 4.68 & 4.20 & 4.60 & 4.30 & 0.1152\tabularnewline
0.24 & 0.0270 & 3.55 & 3.50 & 3.60 & 3.80 & 0.1258\tabularnewline
0.25 & 0.0300 & 3.49 & 3.40 & 3.40 & 3.30 & 0.1333\tabularnewline
0.26 & 0.0213 & 2.51 & 2.80 & 2.80 & 3.00 & 0.1353\tabularnewline
0.27 & 0.0223 & 2.77 & 2.40 & 2.60 & 2.80 & 0.1435\tabularnewline
0.28 & 0.0219 & 2.32 & 1.90 & 2.30 & 2.70 & 0.1405\tabularnewline
0.29 & 0.0077 & - & 1.44 & - & 1.40 & 0.1529\tabularnewline
0.30 & 0.0013 & - & $\approx.35$ & - & - & 0.1598\tabularnewline
0.40 & 0.0011 & - & - & - & - & 0.1523\tabularnewline
0.50 & 0.0001 & - & - & - & - & 0.1490\tabularnewline
\end{tabular}
\par\end{centering}

\end{ruledtabular}
\end{table}

\begin{figure}
\begin{centering}
\includegraphics[scale=0.5]{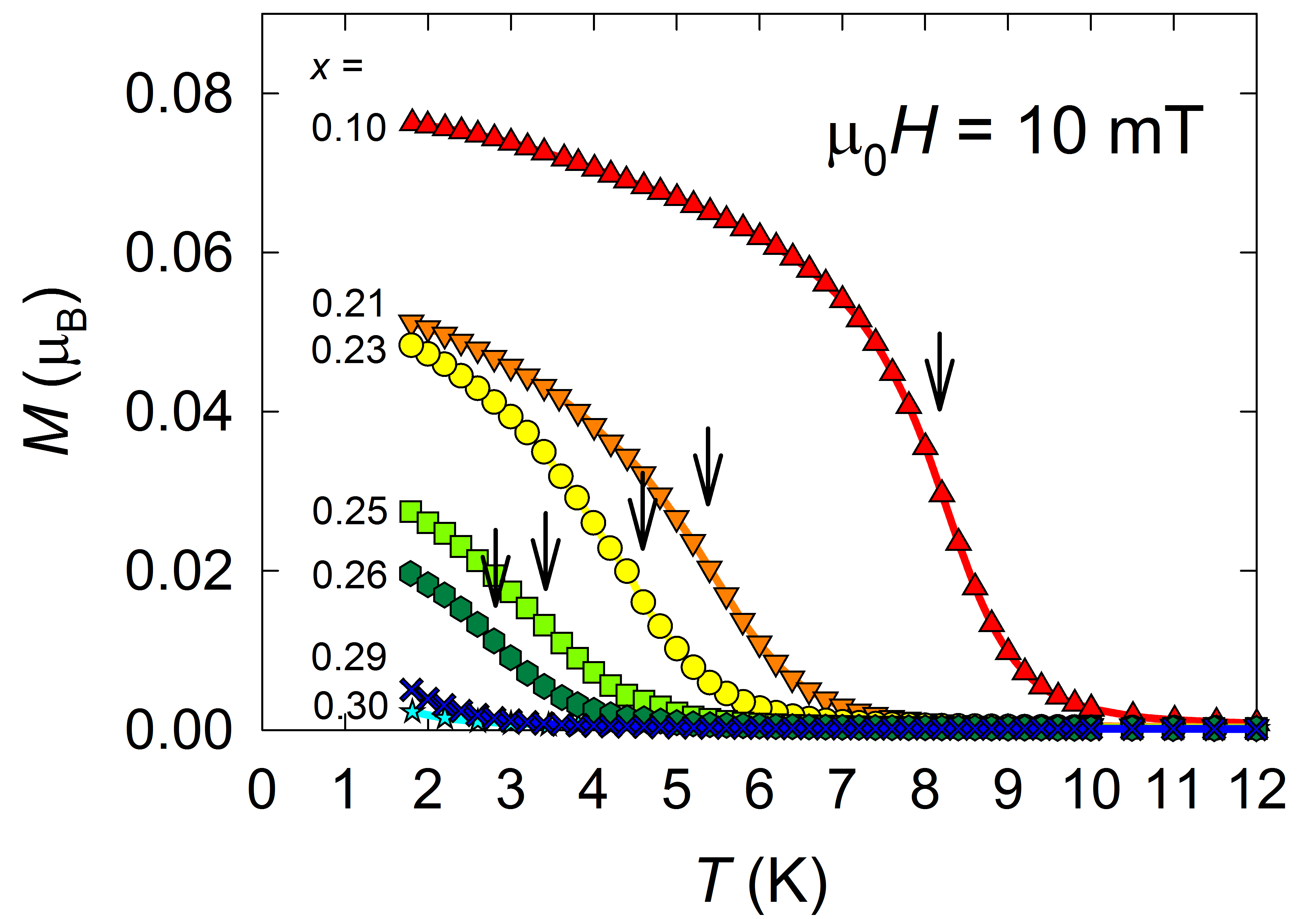}
\par\end{centering}

\protect\caption{\label{fig:(Color-online)Temperature-depend}(Color online) - Temperature
dependence of the magnetization of selected $\mathrm{UCo}_{1-x}\mathrm{Ru}_{x}\mathrm{Ge}$
compounds measured in an external magnetic field of $\unit[10]{mT}$.
The arrows mark $T_{\mathrm{C}}$ for each composition.}
\end{figure}

The $M(T)$ curves measured on selected samples with concentration
above $x\geq0.1$ displayed in Fig. \ref{fig:(Color-online)Temperature-depend}
also manifest the collapse of ferromagnetism with increasing Ru content.
The estimated $T_{\mathrm{C}}$ values as derived from the temperature
of the inflection point in the $M(T)$ dependence (measured in low
external field of $\unit[10]{mT}$) are in good agreement with ordering
temperatures obtained from the Arrott plot analysis (see Table \ref{tab:Values-of-the}
and Fig. \ref{fig:(Color-online)-Panel} (a)). 

\begin{figure}
\begin{centering}
\includegraphics[scale=0.5]{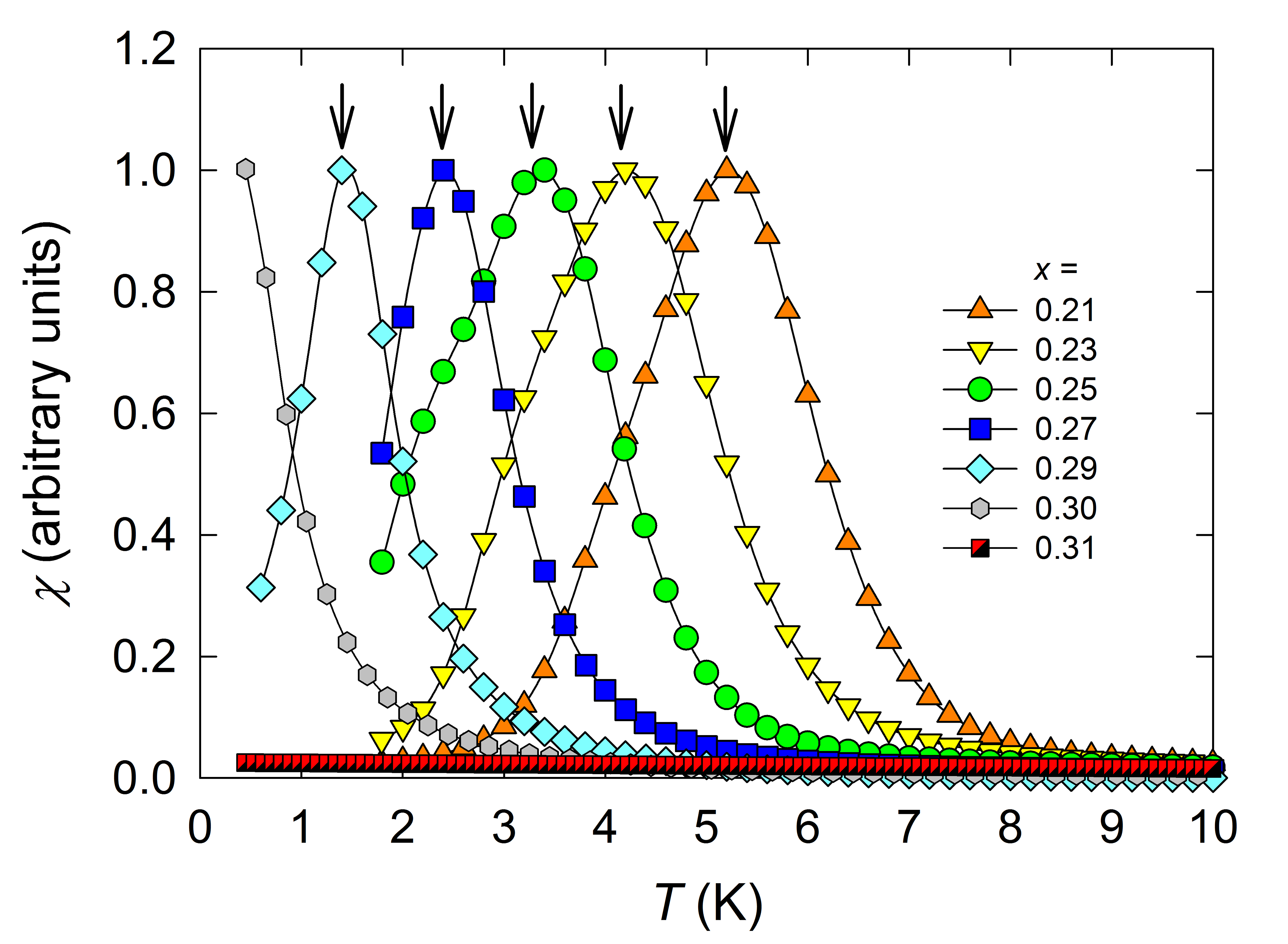}
\par\end{centering}

\protect\caption{\label{fig:(Color-online)-Real}(Color online) - Temperature dependence
of the real part of the AC susceptibility of selected $\mathrm{UCo}_{1-x}\mathrm{Ru}_{x}\mathrm{Ge}$
compounds. The arrows mark $T_{\mathrm{C}}$ for each composition.
Data are plotted in arbitrary units and normalized because the home
made coil for measurement in $^{3}\mathrm{He}$ (used for measurement
of samples with $x=0.29-0.31$) provides only relative data. Some
curves are not shown for clarity of the figure.}

\end{figure}

We have also measured the AC magnetic susceptibility ($\chi$) for
different Ru concentration above $x\geq0.21$ at temperatures down
to $\unit[1.85]{K}$ using a MPMS device. For measurements at lower
temperatures (down to $\unit[400]{mK}$) a custom-made coil system
attached to the $^{3}\mathrm{He}$ insert in PPMS and a lock-in amplifier
were utilized (the same setup as that used in Ref.\cite{Prokleska2010}).
$T_{\mathrm{C}}$ is usually identified as the temperature of the
maximum of the real part of $\chi$ (see Fig. \ref{fig:(Color-online)-Real}).
While the low temperature AC susceptibility of the sample with $x=0.29$
reveal a well-developed peak at $\unit[1.44]{K}$ indicating the onset
of ferromagnetism, no clear peak maximum is observed for the sample
with $x=0.30$, which might be at approximately $\unit[350]{mK}$
as the lowest-$T$ point was measured at $\unit[400]{mK}$. For the
sample with $x=0.31$ no trace of $\chi$ anomaly has been detected
down to $\unit[400]{mK}$ which seems to be in the immediate vicinity
of the critical Ru concentration for existence of ferromagnetism in
the $\mathrm{UCo}_{1-x}\mathrm{Ru}_{x}\mathrm{Ge}$ compounds.

\subsection{Specific heat}

To analyze the different contributions to the specific heat we have
subtracted from experimental data the phonon contribution using the
fit of the phonon specific heat as a $C_{\mathrm{ph}}(T)=\beta T^{3}$.
We typically obtain values of $\beta\approx\unit[(0.52-0.56)\cdot10^{-3}]{J\cdot mol^{-1}K^{-4}}$
which correspond to Debye-temperature values of $\unit[151-155]{K}$.
The remaining part of the specific-heat $C$ represents the sum of
the electronic and magnetic contributions $C_{\mathrm{e}}$ and $C_{\mathrm{m}}$,
respectively.

\begin{figure}
\begin{centering}
\includegraphics[scale=0.5]{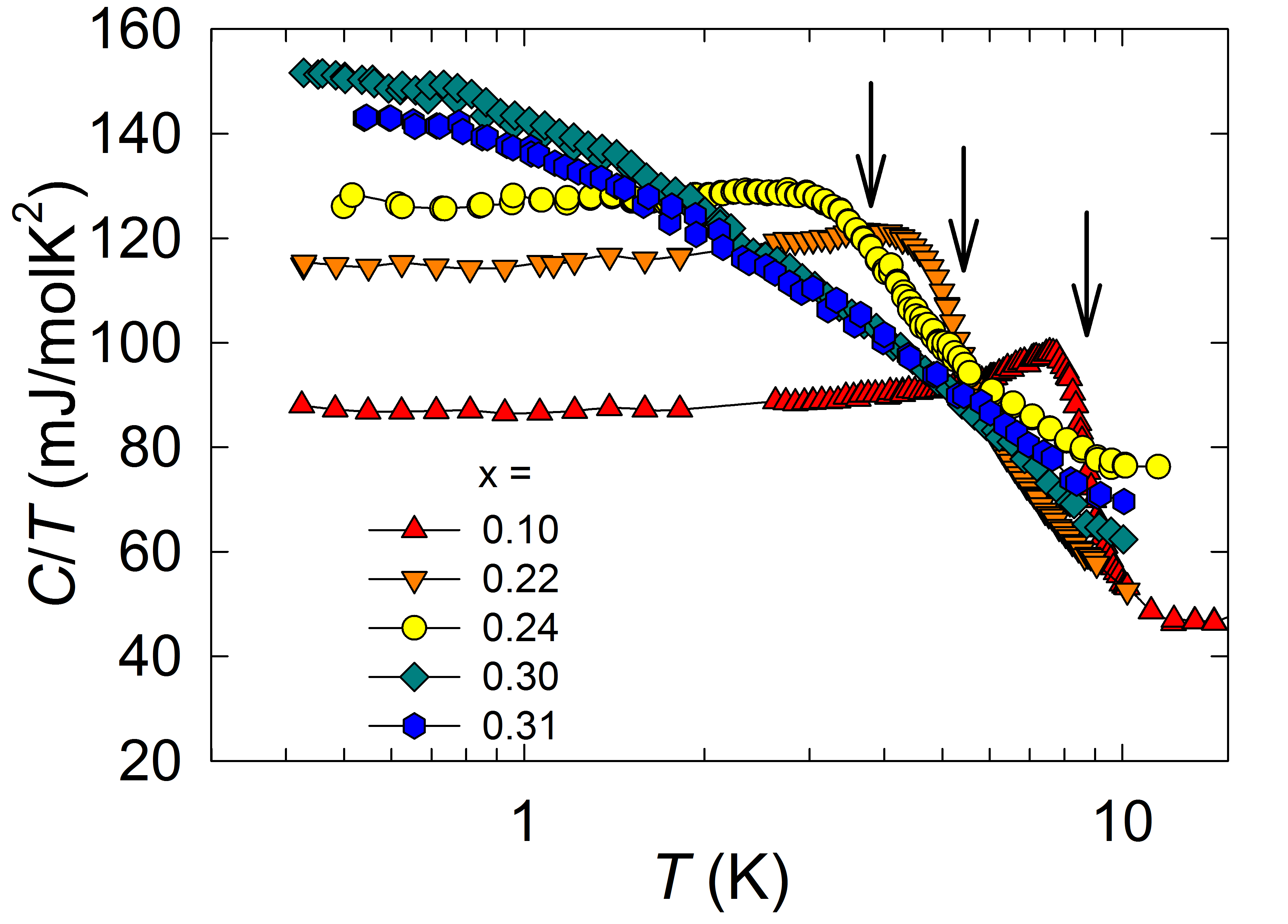}
\par\end{centering}

\protect\caption{\label{fig:(Color-online)---3}(Color online) - $C/T$ versus log$T$
plot for selected $\mathrm{UCo}_{1-x}\mathrm{Ru}_{x}\mathrm{Ge}$
compounds. Black arrows indicate $T_{\mathrm{C}}$ for samples with
$x=0.10$, 0.22 and 0.24, respectively.}

\end{figure}

Fig. \ref{fig:(Color-online)---3} displays the specific heat $C$
divided by temperature $T$ versus $T$ on a log scale for selected
samples between $x=0.1$ and 0.31. The anomaly at $T_{\mathrm{C}}$
is gradually smeared out and shifted to lower temperatures with increasing
Ru concentration. Samples with $x\leq0.3$ show clear anomalies that
are coincident with the onset of ferromagnetic order and are in reasonable
agreement with the $T_{\mathrm{C}}$ values derived from magnetization
and AC susceptibility (see Table \ref{tab:Values-of-the} and Fig.
\ref{fig:(Color-online)-Panel} (a)). For samples with $x=0.30$ and
0.31 $C/T$ versus log$T$ exhibits nearly linear dependence between
1 and $\sim\unit[10]{K}$ but gradually levels off at lower temperatures.
This indicates a non-Fermi-liquid (NFL) behavior $C(T)/T=c\ln\left(T_{0}/T\right)$\cite{Millis1993,Hertz1976}
that is expected for concentrations in the vicinity of the ferromagnetic
QCP. We note that our data do not follow this dependence in the whole
temperature range similar to that recently reported on $\mathrm{UCo}_{1-x}\mathrm{Fe}_{x}\mathrm{Ge}$
system\cite{Huang2013}.

\begin{figure}
\begin{centering}
\includegraphics[scale=0.5]{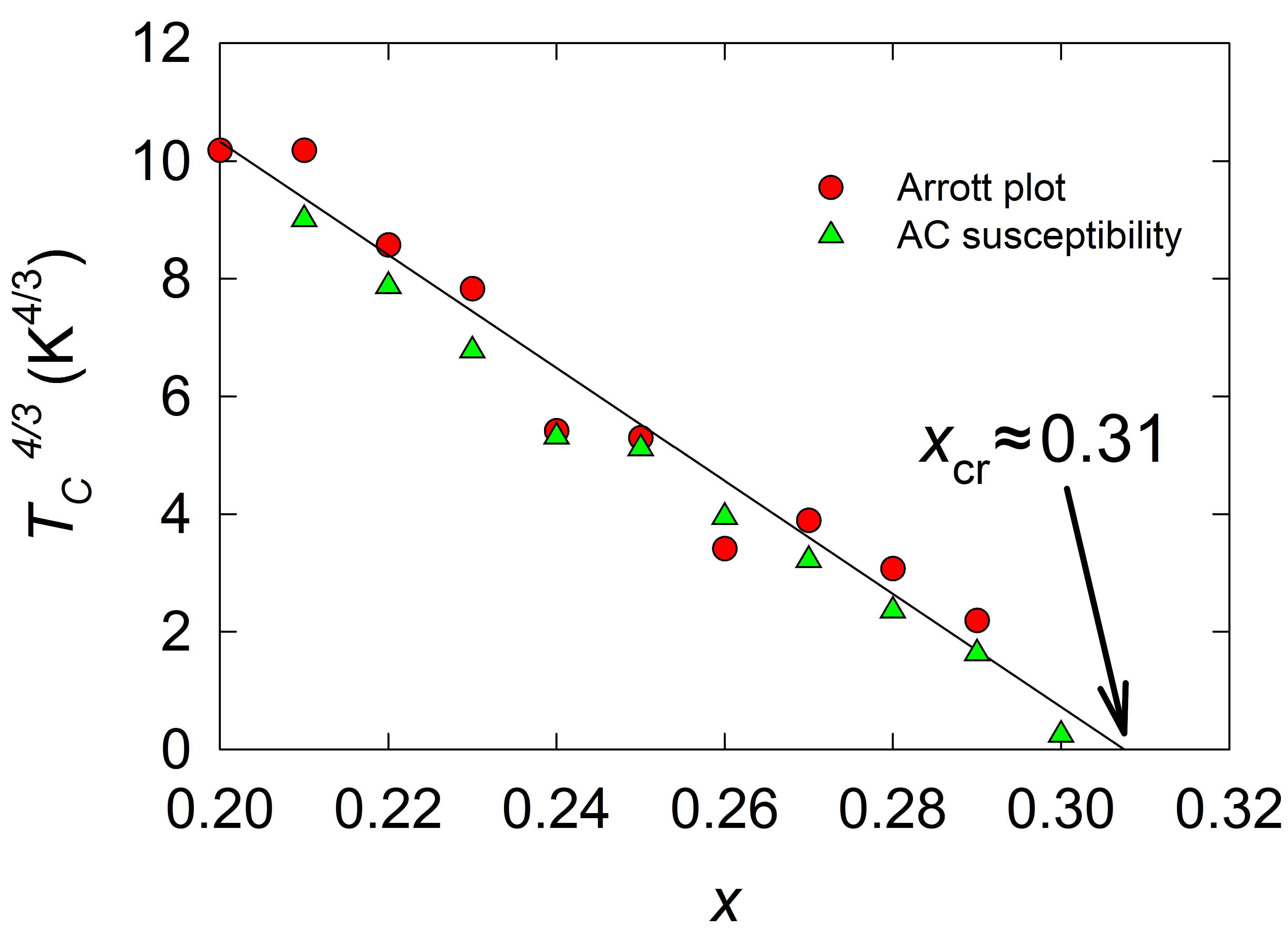}
\par\end{centering}

\protect\caption{\label{fig:(Color-online)-Estimation}(Color online) - Estimation
of the critical concentration for ferromagnetism in the $\mathrm{UCo}_{1-x}\mathrm{Ru}_{x}\mathrm{Ge}$
system by applying the $T_{\mathrm{C}}^{4/3}$ vs $x$ plot and the
$T_{\mathrm{C}}$ values derived from the Arrott plots.}

\end{figure}

We further calculate the magnetic entropy $S_{\mathrm{m}}$ integrated
over the temperature range from $\unit[0.7]{K}$ up to the $T_{\mathrm{C}}$
for each sample and find a steady decrease of $S_{\mathrm{m}}$ with
increasing $x$ from $\unit[0.13]{R\ln2}$ for $x=0.1$ down to $\unit[0.006]{R\ln2}$
at $x=0.30$ (see Fig. \ref{fig:(Color-online)-Panel}(c)). This is
consistent with the observation of a gradual disappearance of the
itinerant magnetic moment by approaching the QCP ($x_{\mathrm{cr}}\approx0.31$).
As the system approaches the critical concentration we observe a large
increase of the value of Sommerfeld coefficient $\gamma$ with a maximum
near $x_{\mathrm{cr}}\approx0.31$ which reflects an enhancement of
the effective mass of the quasiparticles in the region where ferromagnetism
is suppressed. This finding is consistent with the presence of a strong
spin fluctuation near the ferromagnetic QCP. According to the prediction
for the dependence of $T_{\mathrm{C}}$ on a control parameter ($x$)
for itinerant ferromagnets QCP by Millis and Hertz\cite{Millis1993,Hertz1976}
the ordering temperature should obey the relation $T_{\mathrm{C}}\sim\left(x_{\mathrm{cr}}-x\right)^{3/4}$
\cite{Stewart2001} i.e. a linear $T_{\mathrm{C}}^{4/3}$ vs $x$
plot. As we show in Fig. \ref{fig:(Color-online)-Estimation} a linear
fit of $T_{\mathrm{C}}$ values for the samples with $x$ from 0.2
to 0.3 reveals that $T_{\mathrm{C}}$ vanishes at the critical concentration
$x_{\mathrm{cr}}\approx0.31$ consistent with this model. 

\subsection{Electrical resistivity}

\begin{figure}
\begin{centering}
\includegraphics[scale=0.53]{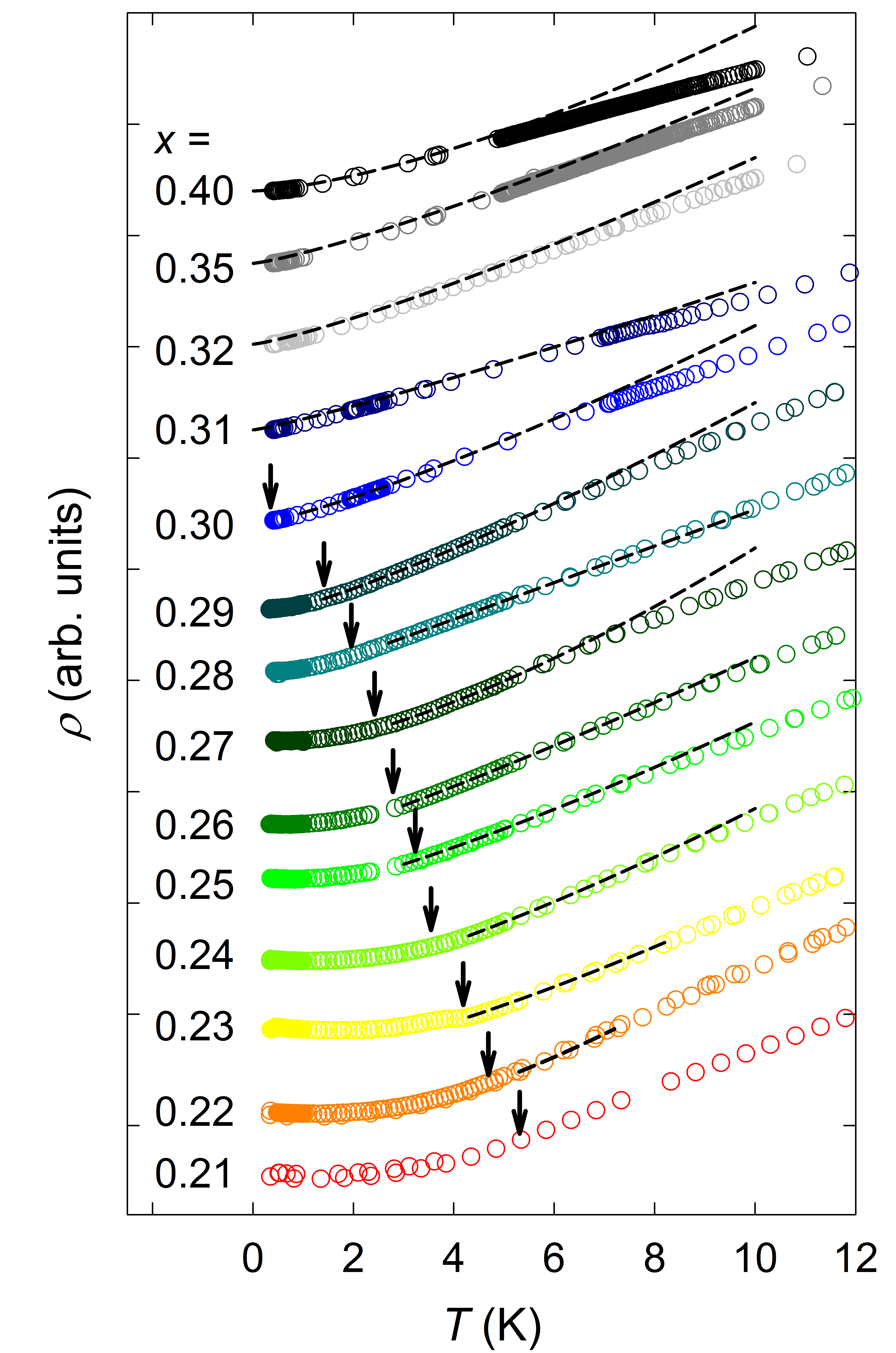}
\par\end{centering}

\protect\caption{\label{fig:(Color-online)-Temperature}(Color online) - Temperature
dependence of the electrical resistivity for selected polycrystalline
samples of $\mathrm{UCo}_{1-x}\mathrm{Ru}_{x}\mathrm{Ge}$. The vertical
arrows denote $T_{\mathrm{C}}$ values obtained from AC susceptibility
data. Dashed lines are fits to data above $T_{\mathrm{C}}$ according
to Eq. (\ref{eq:res}). Each curve is arbitrary vertically shifted
for better clarity of the figure. }

\end{figure}

The low-temperature resistivity data measured on selected polycrystalline
samples are plotted in Fig. \ref{fig:(Color-online)-Temperature}.
Anomalies connected with the transition from paramagnetic to ferromagnetic
state are not clearly visible. It is evident, that increasing Ru content
leads to considerable changes of the low temperature resistivity behavior.
The $\rho\left(T\right)$ data below $T_{\mathrm{C}}$ reasonably
follow the $\rho=\rho_{0}+AT^{2}$ dependence usual for ferromagnets.
Data above $T_{\mathrm{C}}$ were fitted to the relation: 

\begin{equation}
\rho=\rho_{0}+AT^{n}\label{eq:res}
\end{equation}

The inflection point of the $\rho\left(T\right)$ dependence was taken
as an upper limit for the fitting. The exponent ($n$) gradually decreases
as the Ru content approaches the critical concentration $x_{\mathrm{cr}}$.
The minimum value of $n\approx1.13$ for $x=0.31$ is close to the
proposed linear temperature dependence from the theory of Millis and
Hertz\cite{Millis1993,Hertz1976,Stewart2001} for NFL behavior of
a clean 3-dimensional itinerant ferromagnets rather than to the scaling
with the exponent $n=5/3$ which follows from the spin-fluctuation
theory of Moriya\cite{Stewart2001}. The samples with higher concentration
of Ru ($x>x_{\mathrm{cr}}$) seem to exhibit gradual recovery towards
a FL state which is documented by increasing the value of $n$ exponent
with increasing $x$ above $x_{\mathrm{cr}}$.

\begin{figure}
\begin{centering}
\includegraphics[scale=0.45]{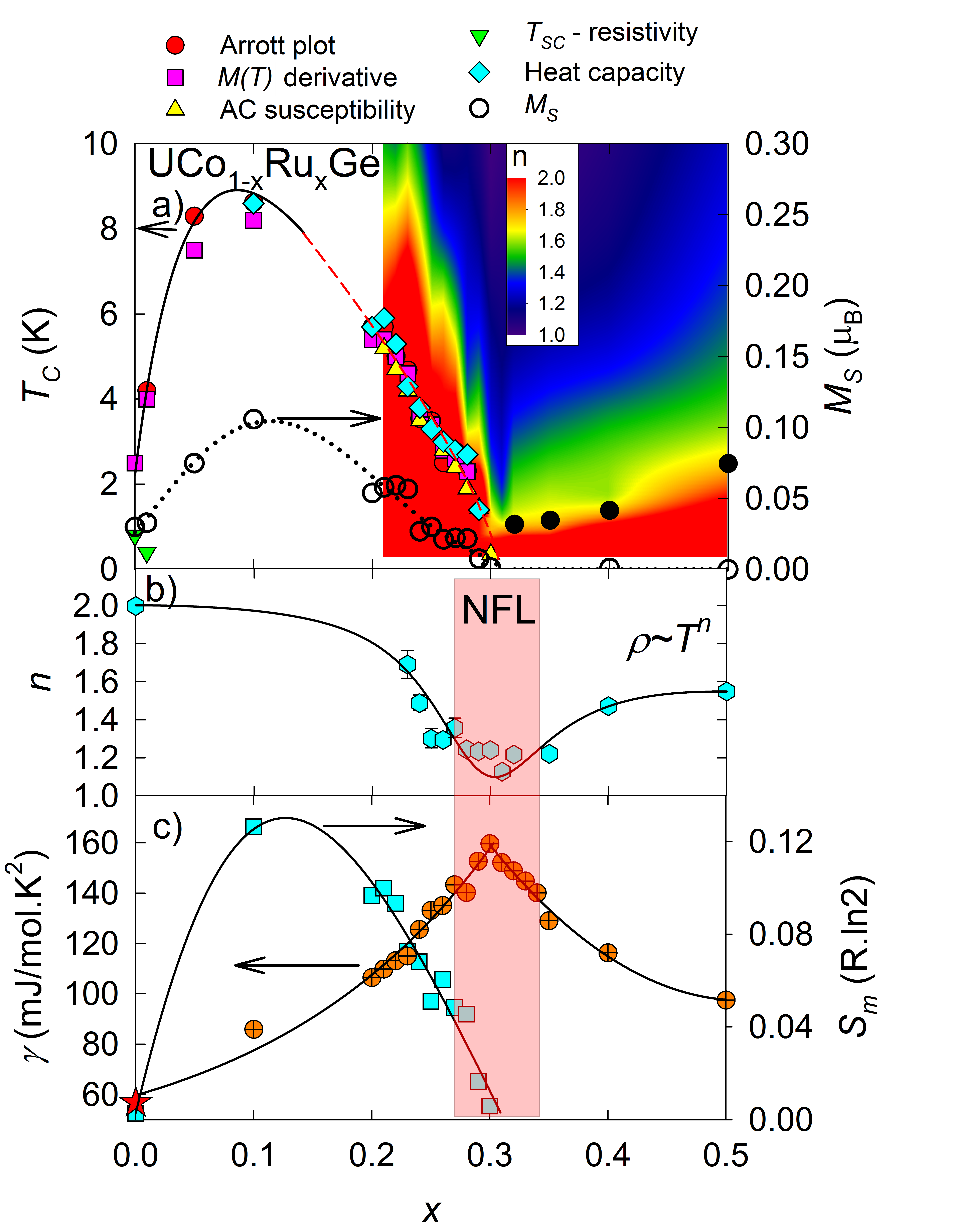}
\par\end{centering}

\protect\caption{\label{fig:(Color-online)-Panel}(Color online) - Panel a) shows the
$T-x$ phase diagram based on measurements of polycrystalline samples.
The diagram is supplemented by the results of the electrical resistivity
measurement revealing occurrence of superconductivity in the parent
UCoGe compound and in $\mathrm{UCo_{0.99}Ru_{0.01}Ge}$ - the two
data points are taken from Ref.\cite{Pospisil2009} (green triangle).
The black solid line is only guide to the eye while the red dashed
part is a fit of $T_{\mathrm{C}}\sim\left(x_{\mathrm{cr}}-x\right)^{3/4}$.
The right axis denotes the spontaneous magnetization $M_{\mathrm{S}}$
(dashed line in the plot is only a guide to the eye). The color plot
shows local exponents of the resistivity obtained as $n=\frac{\mathrm{d}\ln\left(\rho-\rho_{0}\right)}{\mathrm{d\ln T}}$.
The black filled circles show the temperature where resistivity starts
to deviate from the $T^{2}$ dependence. Panel b) shows the evolution
of the coefficients $n$ from the fitting of the low temperature dependence
of the electrical resistivity with equation $\rho=\rho_{0}+AT^{n}$
for $T>T_{\mathrm{C}}$. The right vertical axis shows $RRR=\rho_{\unit[300]{K}}/\rho_{\unit[0.4]{K}}$
as a function of $x$. Panel c) shows development of $C/T$ (extrapolated
to $\unit[0]{K}$) and the magnetic entropy $S_{\mathrm{m}}$ (value
for the parent UCoGe is taken from Ref. \cite{Gasparini2010} and
is marked by a star).}
\end{figure}

Development of the exponent $n$ is summarized in the $T-x$ phase
diagram (Fig. \ref{fig:(Color-online)-Panel} (b)). In order to see
the exponent $n$ as a function of temperature we have calculated
the logarithmic derivative of the electrical resistivity according
to the Eq. (\ref{eq:2-1}).

\begin{equation}
n=\frac{\mathrm{d}\ln\left(\rho-\rho_{0}\right)}{\mathrm{d\ln}T}\label{eq:2-1}
\end{equation}

The results of this analysis are displayed in the colored part of
the phase diagram in Fig. \ref{fig:(Color-online)-Panel} (a). One
can see a significant change of the exponent between the region of
ferromagnetic ordering ($T<T_{\mathrm{C}}$) where $n=2$ and in the
nonmagnetic state where $n<2$. The sharp decrease of the value $n$
near $x_{\mathrm{cr}}$ down to the lowest temperatures is surrounded
by regions of higher $n$ (rapidly increasing on the FM side for $x<x_{\mathrm{cr}}$
and slower increase on the paramagnetic side). 

\subsection{Theoretical calculations}

\begin{figure}
\begin{centering}
\includegraphics[scale=0.5]{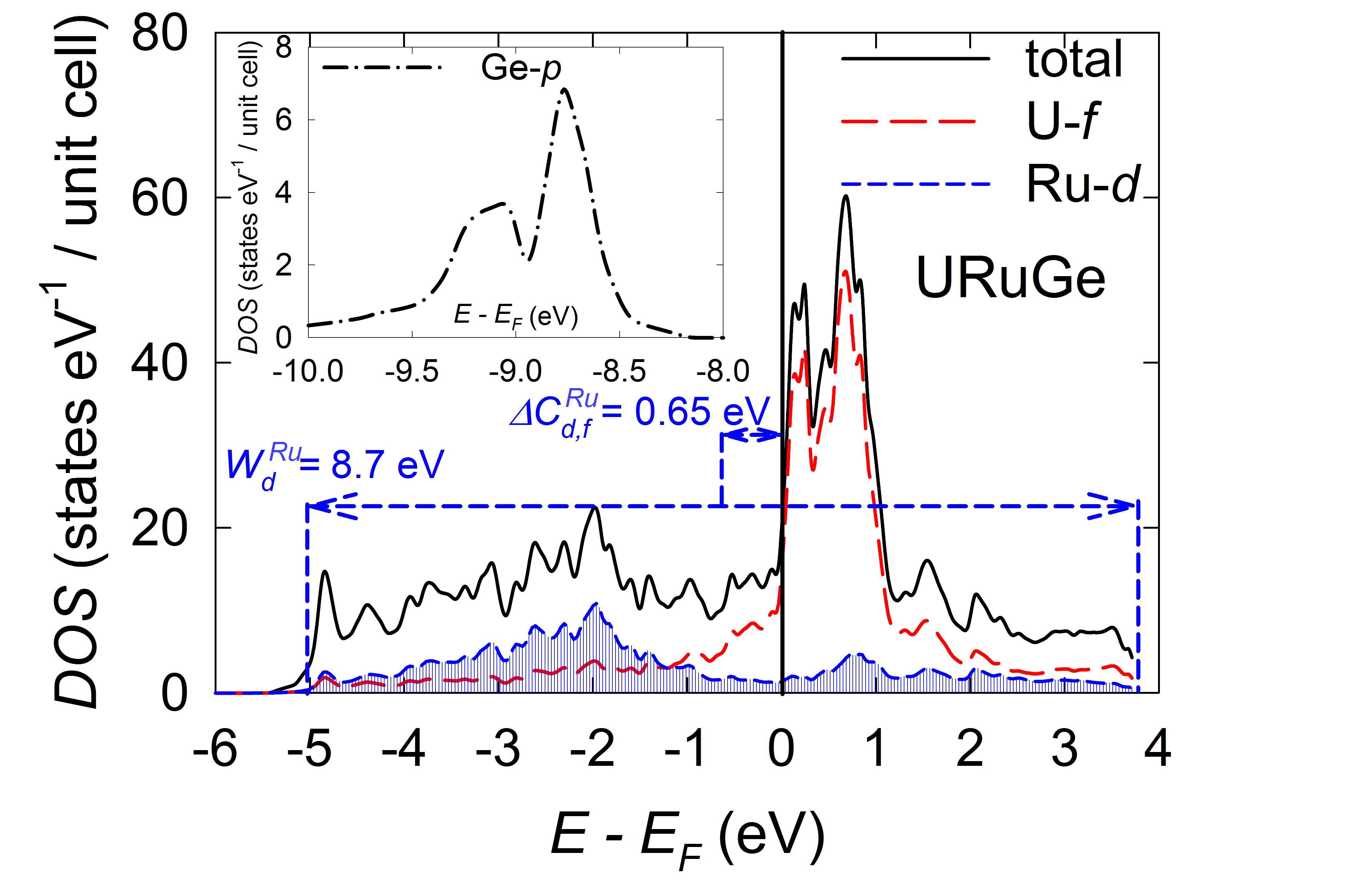}
\par\end{centering}

\protect\caption{\label{fig:(Color-online)-Total}(Color online) - Total and partial
density of states (DOS) for U\textendash $f$ states and Ru\textendash $d$
states in URuGe. Width of the $d$-band ($W_{d}^{Ru}$) and its center
$\Delta C_{df}^{Ru}$ are marked by dashed arrows. Inset shows that
the contribution of the Ge\textendash $p$ states is far from the
Fermi level.}

\end{figure}

In order to better understand the changes in the electronic structure
of the $\mathrm{UCo}_{1-x}\mathrm{Ru}_{x}\mathrm{Ge}$ compounds across
the ferromagnetic QCP, we have performed first-principles theoretical
calculations on the paramagnetic compound URuGe. As a matter of fact,
while the density of states (DOS) of the parent compound UCoGe is
known (Ref. \cite{Divis2008}) the information about the DOS of URuGe
are missing. The calculated total and partial DOS of the URuGe are
plotted in Fig. \ref{fig:(Color-online)-Total}.

We used the calculated URuGe band structure by considering the simple
model of Silva Neto et al. \cite{Silva2013} which is based on the
periodic Anderson model\cite{Batista2002,Batista2003}. This simplified
model proposes the key role of the $nd-5f$ hybridization ($V_{df}$
in Eq. (\ref{eq:3})), where $n$ is the number of $d$ electrons
in the observed non-monotonous evolution of $T_{\mathrm{C}}$ in the
$\mathrm{URh}_{1-x}\mathrm{Co}_{x}$ system. They described the evolution
of $T_{\mathrm{C}}$ with increasing $x$ as a consequence of the
broadening of the $nd$ and $5f$ bands ($W_{d}$, $W_{f}$), respectively,
and the mutual shift of their centers ($C_{Td}-C_{Uf}$) that are
related as \cite{Silva2013}:

\begin{equation}
V_{df}=\frac{W_{d}W_{f}}{C_{Td}-C_{Uf}}\label{eq:3}
\end{equation}

If we apply this model to our $\mathrm{UCo}_{1-x}\mathrm{Ru}_{x}\mathrm{Ge}$
system we can qualitatively describe the non-monotonous evolution
of $T_{\mathrm{C}}$ with Ru concentration. The concentration dependence
of the broadness of the $nd$ band is assumed to be linear according
to \ref{eq:4}

\begin{equation}
W_{d}\left(x\right)=W_{d}^{Co}\left(1-x\right)+W_{d}^{Ru}\left(x\right)\label{eq:4}
\end{equation}

where $W_{d}^{Co}=\unit[6.1]{eV}$(Ref. \cite{Divis2008}) and $W_{d}^{Ru}=\unit[8.7]{eV}$
(see Fig. \ref{fig:(Color-online)-Total}) and $W_{f}=\unit[0.43]{eV}$
(Ref. \cite{Divis2008}). Such a behavior is consistent with other
U$TX$ ($T$= transition metal, $X=$ $p$ element) compounds where
the $d$ band broadens while we move from the 3$d$ to the 4$d$ transition
metals\cite{Gasche1995}. Consequently $\left(C_{Td}-C_{Uf}\right)\left(x\right)=\bigtriangleup C_{df}\left(x\right)$
deviates from linearity

\begin{eqnarray}
\Delta C_{df}(x) & = & \Delta C_{df}^{Co}\left(1-x\right)+\Delta C_{df}^{Ru}(x)+\label{eq:5}\\
 &  & \delta^{\text{\textasciiacute}}x^{2}\left(1-x\right)+\delta^{\text{\textasciiacute\textasciiacute}}x\left(1-x\right)^{2}\nonumber 
\end{eqnarray}

We used the values from calculated DOSes, i.e. $\Delta C_{df}^{Ru}=\unit[0.65]{eV}$
and $\Delta C_{df}^{Co}=\unit[1.5]{eV}$ (Ref. \cite{Divis2008})
and adjustable parameters were taken as $\delta^{\text{\textasciiacute}}=2\cdot10^{-5}$
and $\delta^{\text{\textasciiacute\textasciiacute}}=2$. Such an approach
leads to a non-monotonous dependence of the $d-f$ hybridization term
$V_{df}$; starting with $V_{df}\left(x=0\right)\approx1.73$ for
UCoGe (in agreement with Ref. \cite{Silva2013}), $V_{df}\left(x=1\right)\approx5.55$
for URuGe and $V_{df}\left(x\approx0.3\right)\approx1.9$ as estimated
for the ferromagnetic QCP\cite{Silva2013}. The overall $V_{df}\left(x\right)$
dependence starts with its decrease and thereby causes an enhancement
of the density of $f$ states at the Fermi level $N_{f}\left(E_{F}\right)$\cite{Newns1987}.
In case of itinerant ferromagnets we can estimate the ordering temperature
as a function of the density of states at the Fermi level $T_{\mathrm{C}}\sim\left(IN\left(E_{\mathrm{F}}\right)-1\right)^{3/4}$
where $I$ is the Stoner integral and $N\left(E_{F}\right)$ is the
total density of states at the Fermi level\cite{Moriya2012}. In this
respect we can attribute the initial increase of $T_{\mathrm{C}}$
to the enhanced $N\left(E_{\mathrm{F}}\right)$. At $x\approx0.07$
the $d-f$ hybridization reaches its minimum value $V_{df}=1.7$ and
starts to increase with increasing $x$. This point qualitatively
corresponds to the position of the maximum $T_{\mathrm{C}}$ in the
experimental data at $x\approx0.1$. As the Ru concentration increases
the $d$-band is shifted closer to the position of the $f$-band and
the hybridization increases and thereby results in a reduction of
the contribution of the $N_{f}\left(E_{F}\right)$ to $N\left(E_{F}\right)$\cite{Newns1987}.
For the reason mentioned above the ordering temperature decreases
and reaches zero near $x_{\mathrm{cr}}\approx0.31$.

\section{Discussion}

The 5$f$ electron magnetism in uranium compounds is controlled by
the degree of overlap of the 5$f$ wave functions of neighboring U
ions and by the hybridization of the U-ion 5$f$-electron states with
states of the ligand valence electrons (5$f$-ligand hybridization).
These two mechanisms cause that the 5$f$-electron orbitals loose
their atomic character which they exhibit in the U free ion. Thus,
the 5$f$-5$f$ overlap and/or strong 5$f$-ligand hybridization lead
to delocalization of the 5$f$-electrons, their participation in metallic
bonding\cite{Smith1983}, and consequently a washout of the U magnetic
moment\cite{Cooper1992}. In addition, the spin-orbit interaction
in the U ion plays an important role in electronic structure. Accordingly,
an orbital magnetic moment antiparallel to the spin moment is induced
by the strong spin orbit coupling in the spin-polarized energy bands
of itinerant 5$f$ electron materials\cite{Brooks1983,Lander1991}.
The magnitude of the U 5$f$-electron magnetic moments is thus further
strongly reduced due to the mutual compensation of the orbital and
spin components. The orbital moment is by rule larger than the spin
moment considering results of so far done relevant experiments (see
relevant references in Ref.\cite{Sechovsky1998}).

On the other hand, the 5$f$-ligand hybridization plays a dual role
in U compounds. Besides washing out the 5$f$-electron magnetic moment
it mediates an indirect exchange interaction which couples the uranium
magnetic moments to promote the magnetic ordering and simultaneously
causes very strong magnetocrystalline anisotropy even in very weak
itinerant magnets\cite{Sechovsky1998,Sanchez1995}. Within this process
the hybridized ligand valence states become polarized and as a result
the ligand ion (especially transition element ion) exhibits a small
induced magnetic moment which is usually parallel to the dominant
5$f$-electron orbital component (see relevant references in Ref.\cite{Sechovsky1998}).
This scenario apparently holds for UCoGe as evidenced from a recent
X-ray magnetic circular dichroism (XMCD) study\cite{Taupin2015} and
polarized neutron diffraction (PND) experiments on $\mathrm{UCo_{0.97}Ru_{0.03}Ge}$
and $\mathrm{UCo_{0.88}Ru_{0.12}Ge}$ single crystals\cite{Valiska2015}.
These experiments confirm that the 5$f$-electron orbital moment dominates
the antiparallel spin component. A much smaller Co magnetic moment
is induced by the 5$f$-3$d$ hybridization. 

\begin{figure}
\begin{centering}
\includegraphics[scale=0.4]{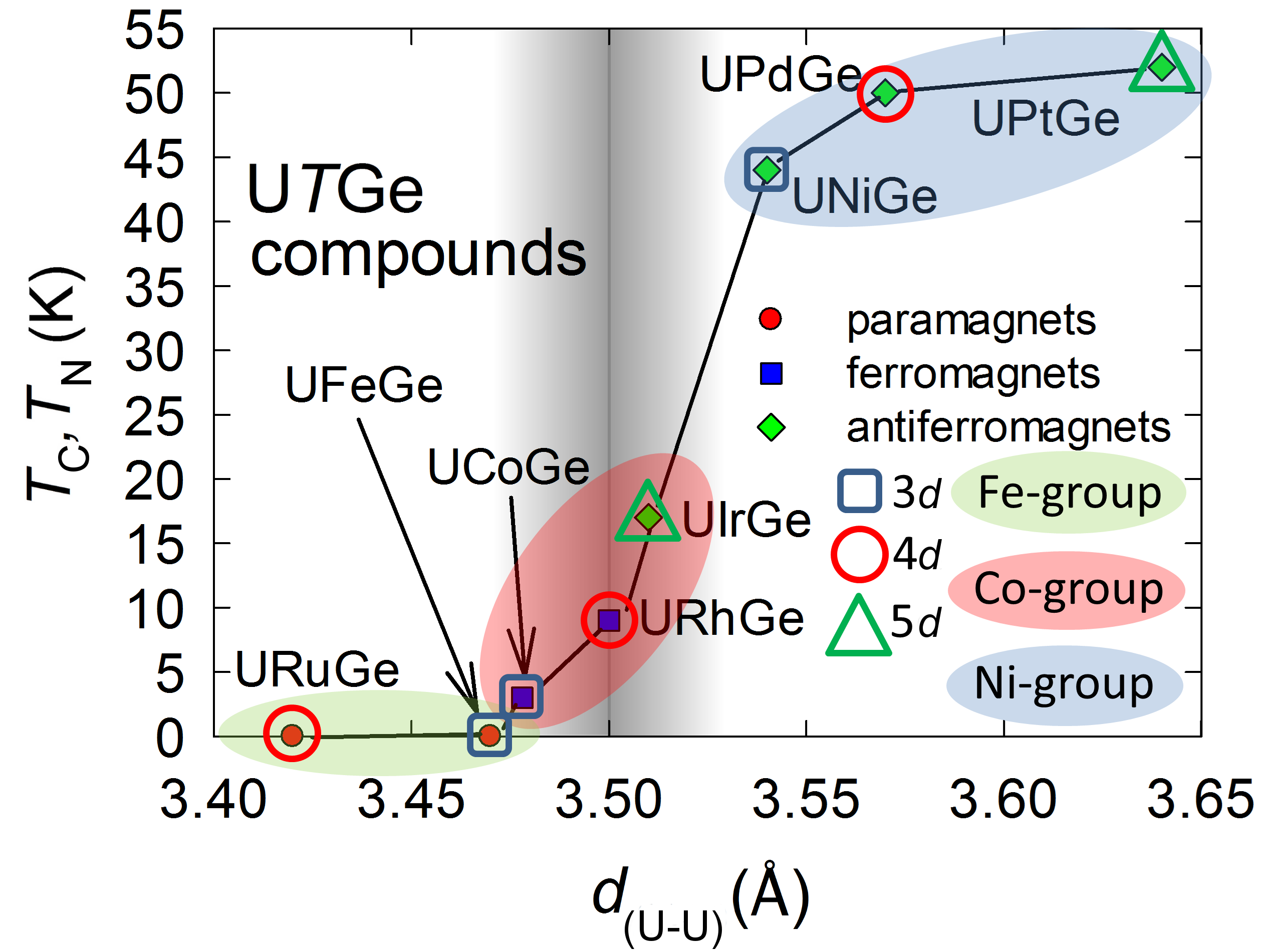}
\par\end{centering}

\protect\caption{\label{fig:(Color-online)-Illustrative}(Color online) - Illustrative
plot showing the dependence of the ordering temperature of the U$T$Ge
compounds ($T$= transition metal) on the shortest distance between
two nearest uranium atoms ($d_{\mathrm{U-U}}$). Shaded region spreads
around Hill limit ($\unit[3.5]{\textrm{\AA}}$)\cite{Hill1970} valid
for uranium. Position of UFeGe is exceptional because UFeGe does not
keep the TiNiSi-type structure \cite{Canepa1996}.}
\end{figure}

Considering the change of the U-U distance $d_{\mathrm{U-U}}$ between
the nearest U neighbor ions (overlap of 5$f$ orbitals) within the
$\mathrm{UCo}_{1-x}\mathrm{Ru}_{x}\mathrm{Ge}$ series, we find that
$d_{\mathrm{U-U}}$ decreases with increasing Ru concentration from
$\unit[\approx3.48]{\textrm{\AA}}$ in UCoGe to $\unit[3.44]{\textrm{\AA}}$
in URuGe (see Fig. \ref{fig:(Color-online)--} and Fig. \ref{fig:(Color-online)-Illustrative}).
Both values fall rather on the \textquotedblleft nonmagnetic side\textquotedblright{}
of the Hill plot\cite{Hill1970}. On the other hand one should bear
in mind that each U ion has only 2 nearest U neighbors on the $d_{\mathrm{U-U}}$
chain meandering along the $a$-axis. If the 5$f$-5$f$ overlap was
the only mechanism controlling magnetism then a gradual washout of
U magnetic moment and monotonously decreasing of $T_{\mathrm{C}}$
with increasing Ru content would be expected. On the contrary, however,
we observe an initial rapid increase of $T_{\mathrm{C}}$ to a maximum
followed by a suppression of ferromagnetism with further increasing
$x$. We note that our observation of a ferromagnetic dome in magnetic
phase diagram in $\mathrm{UCo}_{1-x}\mathrm{Ru}_{x}\mathrm{Ge}$ (see
Fig. \ref{fig:(Color-online)-Panel} (a)) is similar to those observed
for $\mathrm{UCo}_{1-x}\mathrm{Fe}_{x}\mathrm{Ge}$\cite{Huang2013},
$\mathrm{URh}_{1-x}\mathrm{Ru}_{x}\mathrm{Ge}$\cite{Huy2007b} and
$\mathrm{URh}_{1-x}\mathrm{Co}_{x}\mathrm{Ge}$\cite{Huy2009}. 

Apparently an additional mechanism, namely the 5$f$-ligand hybridization
must be taken into account for conceiving the complex evolution of
ferromagnetism in these systems. The increase of $T_{\mathrm{C}}$
and U magnetic moment with increasing $x$ up to 0.12 is accompanied
by increasing the 5$f$ electron orbital moment\cite{Valiska2015}.
The increase of the orbital moment is usually considered as a sign
of partial localization of 5$f$ electrons because the orbital moment
density is distributed closer to the nucleus than the spin density
as it has been demonstrated on a detailed study of the U 5$f$ electron
form factor in $\mathrm{UFe_{2}}$\cite{Lander1991,Wulff1989}. Nevertheless,
the $\mu_{L}/\mu_{S}$ ratio of $\approx2.3$ indicates still a significant
delocalization of the 5$f$ electron states for $x=0.12$\cite{Valiska2015}.
As we mention above our theoretical band structure calculation provide
the basis for understanding the mechanism responsible for the ferromagnetic
dome in the magnetic phase diagram of $\mathrm{UCo}_{1-x}\mathrm{Ru}_{x}\mathrm{Ge}$
by following the simple model treating the changes of 5$f-nd$ hybridization
with variations of the widths and mutual positions on the energy scale
of the transition metal $d$ bands and U 5$f$ bands\cite{Silva2013}.
Accordingly, the non-isoelectronic substitution of Co by Ru causes
broadening of the $d$ band from 3$d$ to the 4$d$ transition metal-like.
Together with the mutual movement of the $d$ and $f$ bands on energy
scale itself we can qualitatively conceive the dome-like dependence
of the ordering temperature $T_{\mathrm{C}}$. This is an important
confirmation of the trend. Variations of the 5$f-nd$ hybridization
most likely cause analogous non-monotonous variation of the magnetic
ground state of $\mathrm{UCo}_{1-x}\mathrm{Fe}_{x}\mathrm{Ge}$\cite{Huang2013}
and $\mathrm{URh}_{1-x}\mathrm{Ru}_{x}\mathrm{Ge}$\cite{Huy2007b}
exhibiting also a ferromagnetic dome in the magnetic phase diagram.
It is worth to mention that the non-monotonous evolution of magnetic
ground state causing a ferromagnetic dome in the magnetic phase diagram
is not only specific to the U$T$Ge compounds possessing the orthorhombic
TiNiSi-type structure. Analogous trends reflecting the varying 5$f-nd$
hybridization are observed also in U$TX$ compounds with the hexagonal
ZrNiAl-type structure. Here UFeAl\cite{Troc1993}, URuAl\cite{Veenhuizen1988}
and UCoAl\cite{Eriksson1989} are paramagnets. The latter compound
is, however, close to a ferromagnetic instability. A magnetic field
of only $\unit[0.6]{T}$ induces in UCoAl itinerant electron metamagnetism\cite{Shimizu1982,Mushnikov1999}.
URhAl\cite{Veenhuizen1988} and URhGe\cite{Sechovsky1998} are ferromagnets.
Ferromagnetic domes are observed in the magnetic phase diagrams of
$\mathrm{UCo}_{1-x}\mathrm{Ru}_{x}\mathrm{Al}$\cite{Andreev1997a},
$\mathrm{URh}_{1-x}\mathrm{Ru}_{x}\mathrm{Al}$\cite{Sechovsky1992},
$\mathrm{URh}_{1-x}\mathrm{Ru}_{x}\mathrm{Ga}$\cite{Sechovsky1992}
and anticipated from the results reported on $\mathrm{UCo}_{1-x}\mathrm{Fe}_{x}\mathrm{Al}$\cite{Mushnikov2002}.

The observed strong delocalization of the 5$f$ electrons in $\mathrm{UCo}_{1-x}\mathrm{Ru}_{x}\mathrm{Ge}$
at higher Ru concentrations is reflected by a dramatic decrease of
the magnetic entropy $S_{\mathrm{m}}$ down to the $\unit[0.006]{R\ln2}$
for $x=0.30$ which points to the itinerant nature of the weak ferromagnetism
in the vicinity of the critical concentration. Note that a magnetic
entropy equal to zero is expected for an ideal itinerant electron
ferromagnet\cite{Sechovsky1998}. Our results of the temperature dependence
of the electrical resistivity provide evidence for a NFL behavior
in the vicinity of $x_{\mathrm{cr}}$ most likely caused by the possible
presence of the FM QCP. We have observed a drop of the $n$ exponent
in the temperature dependence of resistivity $\rho=\rho_{0}+AT^{n}$
and an almost logarithmic dependence of the heat capacity $C(T)/T=c\ln\left(T_{0}/T\right)$
in a limited interval at lowest temperatures that would be in agreement
with the theoretical predictions of Millis and Hertz\cite{Millis1993,Hertz1976}.
Further evidence for the FM QCP is offered by the rapid increase of
the effective mass of the quasiparticles near $x_{\mathrm{cr}}$.
The proposed scenario is also corroborated by scaling of the ordering
temperature with the control parameter itself which obeys the formula
$T_{\mathrm{C}}\sim\left(x_{\mathrm{cr}}-x\right)^{3/4}$ and provides
estimation of the critical concentration $x_{\mathrm{cr}}\approx0.31$.

The FM transition of $\mathrm{UCo}_{1-x}\mathrm{Ru}_{x}\mathrm{Ge}$
compounds in the vicinity of $x_{\mathrm{cr}}$ is apparently of a
second order type in contrast to the first order transition reported
for 3-dimensional ferromagnets in the vicinity of a QCP\cite{Belitz2012}.
Microscopic NQR studies of UCoGe suggest a first order transition
to the FM state\cite{Hattori2010}. The second order transition in
$\mathrm{UCo}_{1-x}\mathrm{Ru}_{x}\mathrm{Ge}$ compounds near $x_{\mathrm{cr}}$
can be conceived as a consequence of the substitution- induced disorder
in the system which may blur the first order transition towards a
continuous second order transition. In this context, we would like
to mention the experimental and theoretical arguments regarding the
observed anomalies related to the existence of a ferromagnetic QCP
in $\mathrm{UCo}_{1-x}\mathrm{Ru}_{x}\mathrm{Ge}$ should be considered
with a proper caution. Disorder caused by substitution can in some
cases emulate NFL behavior\cite{Miranda1997,Rosch1999} and may be
one of the reasons of the lacking superconductivity in $\mathrm{UCo}_{1-x}\mathrm{Ru}_{x}\mathrm{Ge}$
in the proximity of the QCP. Thorough the investigation of single
crystals of $\mathrm{UCo}_{1-x}\mathrm{Ru}_{x}\mathrm{Ge}$ compounds
near $x_{\mathrm{cr}}$ at ambient and high pressures is highly desired
in order to clarify the origin of the NFL state and the character
of the ferromagnetic quantum phase transition in $\mathrm{UCo}_{1-x}\mathrm{Ru}_{x}\mathrm{Ge}$.

\section{Conclusions}

We have successfully prepared series of the polycrystalline samples
of UCoGe doped with Ru in a wide range of concentration $\mathrm{UCo}_{1-x}\mathrm{Ru}_{x}\mathrm{Ge}$
($0\geq x\leq0.9$). The Ru substitution leads to development of a
FM dome between $x=0-0.31$ with the maximum of $T_{\mathrm{C}}=\unit[8.6]{K}$
and $M_{\mathrm{S}}=\unit[0.1]{\mu_{\mathrm{B}}}$ appearing at the
$x\approx0.1$. Further increase of the Ru content up to the critical
concentration $x_{\mathrm{cr}}\approx0.31$ leads to the disappearance
of the ferromagnetic state at a QCP. Using electronic structure calculations
we were able to explain the evolution of ferromagnetism with $x$
for $\mathrm{UCo}_{1-x}\mathrm{Ru}_{x}\mathrm{Ge}$ in terms of changes
of the density of states at the Fermi level due to varying 5$f$-ligand
hybridization. The analysis of the critical exponents of the electrical
resistivity and heat capacity at low temperatures revealed a non-Fermi
liquid behavior for the samples in the vicinity of the QCP. The NFL
state can be influenced by the substitution-induced disorder of the
system because of the non-isoelectronic mixture of the 3$d$ (Co)
and 4$d$ (Ru) bands. Further study of the region around the critical
concentration including the measurements under the external pressure
performed on high quality single crystals is highly desired for a
better understanding the physics underlying the ferromagnetic quantum
phase transition. 
\begin{acknowledgments}
This work was supported by the Czech Science Foundation no. P204/12/P418
and the Charles University in Prague, project GA UK No.720214. Experiments
performed in MLTL (see: http://mltl.eu/) were supported within the
program of Czech Research Infrastructures (project LM2011025).
\end{acknowledgments}

\bibliographystyle{apsrev4-1}
\bibliography{UCoGe}

\begin{thebibliography}{72}%
\makeatletter
\providecommand \@ifxundefined [1]{%
 \@ifx{#1\undefined}
}%
\providecommand \@ifnum [1]{%
 \ifnum #1\expandafter \@firstoftwo
 \else \expandafter \@secondoftwo
 \fi
}%
\providecommand \@ifx [1]{%
 \ifx #1\expandafter \@firstoftwo
 \else \expandafter \@secondoftwo
 \fi
}%
\providecommand \natexlab [1]{#1}%
\providecommand \enquote  [1]{``#1''}%
\providecommand \bibnamefont  [1]{#1}%
\providecommand \bibfnamefont [1]{#1}%
\providecommand \citenamefont [1]{#1}%
\providecommand \href@noop [0]{\@secondoftwo}%
\providecommand \href [0]{\begingroup \@sanitize@url \@href}%
\providecommand \@href[1]{\@@startlink{#1}\@@href}%
\providecommand \@@href[1]{\endgroup#1\@@endlink}%
\providecommand \@sanitize@url [0]{\catcode `\\12\catcode `\$12\catcode
  `\&12\catcode `\#12\catcode `\^12\catcode `\_12\catcode `\%12\relax}%
\providecommand \@@startlink[1]{}%
\providecommand \@@endlink[0]{}%
\providecommand \url  [0]{\begingroup\@sanitize@url \@url }%
\providecommand \@url [1]{\endgroup\@href {#1}{\urlprefix }}%
\providecommand \urlprefix  [0]{URL }%
\providecommand \Eprint [0]{\href }%
\providecommand \doibase [0]{http://dx.doi.org/}%
\providecommand \selectlanguage [0]{\@gobble}%
\providecommand \bibinfo  [0]{\@secondoftwo}%
\providecommand \bibfield  [0]{\@secondoftwo}%
\providecommand \translation [1]{[#1]}%
\providecommand \BibitemOpen [0]{}%
\providecommand \bibitemStop [0]{}%
\providecommand \bibitemNoStop [0]{.\EOS\space}%
\providecommand \EOS [0]{\spacefactor3000\relax}%
\providecommand \BibitemShut  [1]{\csname bibitem#1\endcsname}%
\let\auto@bib@innerbib\@empty
\bibitem [{\citenamefont {Saxena}\ \emph {et~al.}(2000)\citenamefont {Saxena},
  \citenamefont {Agarwal}, \citenamefont {Ahilan}, \citenamefont {Grosche},
  \citenamefont {Haselwimmer}, \citenamefont {Steiner}, \citenamefont {Pugh},
  \citenamefont {Walker}, \citenamefont {Julian}, \citenamefont {Monthoux},
  \citenamefont {Lonzarich}, \citenamefont {Huxley}, \citenamefont {Sheikin},
  \citenamefont {Braithwaite},\ and\ \citenamefont {Flouquet}}]{Saxena2000}%
  \BibitemOpen
  \bibfield  {author} {\bibinfo {author} {\bibfnamefont {S.~S.}\ \bibnamefont
  {Saxena}}, \bibinfo {author} {\bibfnamefont {P.}~\bibnamefont {Agarwal}},
  \bibinfo {author} {\bibfnamefont {K.}~\bibnamefont {Ahilan}}, \bibinfo
  {author} {\bibfnamefont {F.~M.}\ \bibnamefont {Grosche}}, \bibinfo {author}
  {\bibfnamefont {R.~K.~W.}\ \bibnamefont {Haselwimmer}}, \bibinfo {author}
  {\bibfnamefont {M.~J.}\ \bibnamefont {Steiner}}, \bibinfo {author}
  {\bibfnamefont {E.}~\bibnamefont {Pugh}}, \bibinfo {author} {\bibfnamefont
  {I.~R.}\ \bibnamefont {Walker}}, \bibinfo {author} {\bibfnamefont {S.~R.}\
  \bibnamefont {Julian}}, \bibinfo {author} {\bibfnamefont {P.}~\bibnamefont
  {Monthoux}}, \bibinfo {author} {\bibfnamefont {G.~G.}\ \bibnamefont
  {Lonzarich}}, \bibinfo {author} {\bibfnamefont {A.}~\bibnamefont {Huxley}},
  \bibinfo {author} {\bibfnamefont {I.}~\bibnamefont {Sheikin}}, \bibinfo
  {author} {\bibfnamefont {D.}~\bibnamefont {Braithwaite}}, \ and\ \bibinfo
  {author} {\bibfnamefont {J.}~\bibnamefont {Flouquet}},\ }\href@noop {}
  {\bibfield  {journal} {\bibinfo  {journal} {Nature}\ }\textbf {\bibinfo
  {volume} {406}},\ \bibinfo {pages} {587} (\bibinfo {year}
  {2000})}\BibitemShut {NoStop}%
\bibitem [{\citenamefont {Huxley}\ \emph {et~al.}(2001)\citenamefont {Huxley},
  \citenamefont {Sheikin}, \citenamefont {Ressouche}, \citenamefont
  {Kernavanois}, \citenamefont {Braithwaite}, \citenamefont {Calemczuk},\ and\
  \citenamefont {Flouquet}}]{Huxley2001}%
  \BibitemOpen
  \bibfield  {author} {\bibinfo {author} {\bibfnamefont {A.}~\bibnamefont
  {Huxley}}, \bibinfo {author} {\bibfnamefont {I.}~\bibnamefont {Sheikin}},
  \bibinfo {author} {\bibfnamefont {E.}~\bibnamefont {Ressouche}}, \bibinfo
  {author} {\bibfnamefont {N.}~\bibnamefont {Kernavanois}}, \bibinfo {author}
  {\bibfnamefont {D.}~\bibnamefont {Braithwaite}}, \bibinfo {author}
  {\bibfnamefont {R.}~\bibnamefont {Calemczuk}}, \ and\ \bibinfo {author}
  {\bibfnamefont {J.}~\bibnamefont {Flouquet}},\ }\href@noop {} {\bibfield
  {journal} {\bibinfo  {journal} {Physical Review B}\ }\textbf {\bibinfo
  {volume} {63}},\ \bibinfo {pages} {144519} (\bibinfo {year}
  {2001})}\BibitemShut {NoStop}%
\bibitem [{\citenamefont {Aoki}\ \emph {et~al.}(2001)\citenamefont {Aoki},
  \citenamefont {Huxley}, \citenamefont {Ressouche}, \citenamefont
  {Braithwaite}, \citenamefont {Flouquet}, \citenamefont {Brison},
  \citenamefont {Lhotel},\ and\ \citenamefont {Paulsen}}]{Aoki2001}%
  \BibitemOpen
  \bibfield  {author} {\bibinfo {author} {\bibfnamefont {D.}~\bibnamefont
  {Aoki}}, \bibinfo {author} {\bibfnamefont {A.}~\bibnamefont {Huxley}},
  \bibinfo {author} {\bibfnamefont {E.}~\bibnamefont {Ressouche}}, \bibinfo
  {author} {\bibfnamefont {D.}~\bibnamefont {Braithwaite}}, \bibinfo {author}
  {\bibfnamefont {J.}~\bibnamefont {Flouquet}}, \bibinfo {author}
  {\bibfnamefont {J.-P.}\ \bibnamefont {Brison}}, \bibinfo {author}
  {\bibfnamefont {E.}~\bibnamefont {Lhotel}}, \ and\ \bibinfo {author}
  {\bibfnamefont {C.}~\bibnamefont {Paulsen}},\ }\href@noop {} {\bibfield
  {journal} {\bibinfo  {journal} {Nature}\ }\textbf {\bibinfo {volume} {413}},\
  \bibinfo {pages} {613} (\bibinfo {year} {2001})}\BibitemShut {NoStop}%
\bibitem [{\citenamefont {Huy}\ \emph {et~al.}(2007{\natexlab{a}})\citenamefont
  {Huy}, \citenamefont {Gasparini}, \citenamefont {de~Nijs}, \citenamefont
  {Huang}, \citenamefont {Klaasse}, \citenamefont {Gortenmulder}, \citenamefont
  {de~Visser}, \citenamefont {Hamann}, \citenamefont {Gorlach},\ and\
  \citenamefont {von Lohneysen}}]{Huy2007}%
  \BibitemOpen
  \bibfield  {author} {\bibinfo {author} {\bibfnamefont {N.~T.}\ \bibnamefont
  {Huy}}, \bibinfo {author} {\bibfnamefont {A.}~\bibnamefont {Gasparini}},
  \bibinfo {author} {\bibfnamefont {D.~E.}\ \bibnamefont {de~Nijs}}, \bibinfo
  {author} {\bibfnamefont {Y.}~\bibnamefont {Huang}}, \bibinfo {author}
  {\bibfnamefont {J.~C.~P.}\ \bibnamefont {Klaasse}}, \bibinfo {author}
  {\bibfnamefont {T.}~\bibnamefont {Gortenmulder}}, \bibinfo {author}
  {\bibfnamefont {A.}~\bibnamefont {de~Visser}}, \bibinfo {author}
  {\bibfnamefont {A.}~\bibnamefont {Hamann}}, \bibinfo {author} {\bibfnamefont
  {T.}~\bibnamefont {Gorlach}}, \ and\ \bibinfo {author} {\bibfnamefont
  {H.}~\bibnamefont {von Lohneysen}},\ }\href@noop {} {\bibfield  {journal}
  {\bibinfo  {journal} {Physical Review Letters}\ }\textbf {\bibinfo {volume}
  {99}},\ \bibinfo {pages} {067006} (\bibinfo {year}
  {2007}{\natexlab{a}})}\BibitemShut {NoStop}%
\bibitem [{\citenamefont {Uhlarz}\ \emph {et~al.}(2004)\citenamefont {Uhlarz},
  \citenamefont {Pfleiderer},\ and\ \citenamefont {Hayden}}]{Uhlarz2004}%
  \BibitemOpen
  \bibfield  {author} {\bibinfo {author} {\bibfnamefont {M.}~\bibnamefont
  {Uhlarz}}, \bibinfo {author} {\bibfnamefont {C.}~\bibnamefont {Pfleiderer}},
  \ and\ \bibinfo {author} {\bibfnamefont {S.~M.}\ \bibnamefont {Hayden}},\
  }\href@noop {} {\bibfield  {journal} {\bibinfo  {journal} {Physical Review
  Letters}\ }\textbf {\bibinfo {volume} {93}},\ \bibinfo {pages} {256404}
  (\bibinfo {year} {2004})}\BibitemShut {NoStop}%
\bibitem [{\citenamefont {de~Nijs}\ \emph {et~al.}(2008)\citenamefont
  {de~Nijs}, \citenamefont {Huy},\ and\ \citenamefont
  {de~Visser}}]{deNijs2008}%
  \BibitemOpen
  \bibfield  {author} {\bibinfo {author} {\bibfnamefont {D.~E.}\ \bibnamefont
  {de~Nijs}}, \bibinfo {author} {\bibfnamefont {N.~T.}\ \bibnamefont {Huy}}, \
  and\ \bibinfo {author} {\bibfnamefont {A.}~\bibnamefont {de~Visser}},\
  }\href@noop {} {\bibfield  {journal} {\bibinfo  {journal} {Physical Review
  B}\ }\textbf {\bibinfo {volume} {77}},\ \bibinfo {pages} {140506R} (\bibinfo
  {year} {2008})}\BibitemShut {NoStop}%
\bibitem [{\citenamefont {Stock}\ \emph {et~al.}(2011)\citenamefont {Stock},
  \citenamefont {Sokolov}, \citenamefont {Bourges}, \citenamefont {Tobash},
  \citenamefont {Gofryk}, \citenamefont {Ronning}, \citenamefont {Bauer},
  \citenamefont {Rule},\ and\ \citenamefont {Huxley}}]{Stock2011}%
  \BibitemOpen
  \bibfield  {author} {\bibinfo {author} {\bibfnamefont {C.}~\bibnamefont
  {Stock}}, \bibinfo {author} {\bibfnamefont {D.~A.}\ \bibnamefont {Sokolov}},
  \bibinfo {author} {\bibfnamefont {P.}~\bibnamefont {Bourges}}, \bibinfo
  {author} {\bibfnamefont {P.~H.}\ \bibnamefont {Tobash}}, \bibinfo {author}
  {\bibfnamefont {K.}~\bibnamefont {Gofryk}}, \bibinfo {author} {\bibfnamefont
  {F.}~\bibnamefont {Ronning}}, \bibinfo {author} {\bibfnamefont {E.~D.}\
  \bibnamefont {Bauer}}, \bibinfo {author} {\bibfnamefont {K.~C.}\ \bibnamefont
  {Rule}}, \ and\ \bibinfo {author} {\bibfnamefont {A.~D.}\ \bibnamefont
  {Huxley}},\ }\href@noop {} {\bibfield  {journal} {\bibinfo  {journal}
  {Physical Review Letters}\ }\textbf {\bibinfo {volume} {107}},\ \bibinfo
  {pages} {187202} (\bibinfo {year} {2011})}\BibitemShut {NoStop}%
\bibitem [{\citenamefont {Hattori}\ \emph {et~al.}(2012)\citenamefont
  {Hattori}, \citenamefont {Ihara}, \citenamefont {Nakai}, \citenamefont
  {Ishida}, \citenamefont {Tada}, \citenamefont {Fujimoto}, \citenamefont
  {Kawakami}, \citenamefont {Osaki}, \citenamefont {Deguchi}, \citenamefont
  {Sato},\ and\ \citenamefont {Satoh}}]{Hatorri2012}%
  \BibitemOpen
  \bibfield  {author} {\bibinfo {author} {\bibfnamefont {T.}~\bibnamefont
  {Hattori}}, \bibinfo {author} {\bibfnamefont {Y.}~\bibnamefont {Ihara}},
  \bibinfo {author} {\bibfnamefont {Y.}~\bibnamefont {Nakai}}, \bibinfo
  {author} {\bibfnamefont {K.}~\bibnamefont {Ishida}}, \bibinfo {author}
  {\bibfnamefont {Y.}~\bibnamefont {Tada}}, \bibinfo {author} {\bibfnamefont
  {S.}~\bibnamefont {Fujimoto}}, \bibinfo {author} {\bibfnamefont
  {N.}~\bibnamefont {Kawakami}}, \bibinfo {author} {\bibfnamefont
  {E.}~\bibnamefont {Osaki}}, \bibinfo {author} {\bibfnamefont
  {K.}~\bibnamefont {Deguchi}}, \bibinfo {author} {\bibfnamefont {N.~K.}\
  \bibnamefont {Sato}}, \ and\ \bibinfo {author} {\bibfnamefont
  {I.}~\bibnamefont {Satoh}},\ }\href@noop {} {\bibfield  {journal} {\bibinfo
  {journal} {Physical Review Letters}\ }\textbf {\bibinfo {volume} {108}}
  (\bibinfo {year} {2012})}\BibitemShut {NoStop}%
\bibitem [{\citenamefont {Marel}\ \emph {et~al.}(2003)\citenamefont {Marel},
  \citenamefont {Molegraaf}, \citenamefont {Zaanen}, \citenamefont {Nussinov},
  \citenamefont {Carbone}, \citenamefont {Damascelli}, \citenamefont {Eisaki},
  \citenamefont {Greven}, \citenamefont {Kes},\ and\ \citenamefont
  {Li}}]{Marel2003}%
  \BibitemOpen
  \bibfield  {author} {\bibinfo {author} {\bibfnamefont {D.~v.~d.}\
  \bibnamefont {Marel}}, \bibinfo {author} {\bibfnamefont {H.~J.~A.}\
  \bibnamefont {Molegraaf}}, \bibinfo {author} {\bibfnamefont {J.}~\bibnamefont
  {Zaanen}}, \bibinfo {author} {\bibfnamefont {Z.}~\bibnamefont {Nussinov}},
  \bibinfo {author} {\bibfnamefont {F.}~\bibnamefont {Carbone}}, \bibinfo
  {author} {\bibfnamefont {A.}~\bibnamefont {Damascelli}}, \bibinfo {author}
  {\bibfnamefont {H.}~\bibnamefont {Eisaki}}, \bibinfo {author} {\bibfnamefont
  {M.}~\bibnamefont {Greven}}, \bibinfo {author} {\bibfnamefont {P.~H.}\
  \bibnamefont {Kes}}, \ and\ \bibinfo {author} {\bibfnamefont
  {M.}~\bibnamefont {Li}},\ }\href@noop {} {\bibfield  {journal} {\bibinfo
  {journal} {Nature}\ }\textbf {\bibinfo {volume} {425}},\ \bibinfo {pages}
  {271} (\bibinfo {year} {2003})}\BibitemShut {NoStop}%
\bibitem [{\citenamefont {Pfleiderer}\ \emph {et~al.}(1997)\citenamefont
  {Pfleiderer}, \citenamefont {McMullan}, \citenamefont {Julian},\ and\
  \citenamefont {Lonzarich}}]{Pfleiderer1997}%
  \BibitemOpen
  \bibfield  {author} {\bibinfo {author} {\bibfnamefont {C.}~\bibnamefont
  {Pfleiderer}}, \bibinfo {author} {\bibfnamefont {G.~J.}\ \bibnamefont
  {McMullan}}, \bibinfo {author} {\bibfnamefont {S.~R.}\ \bibnamefont
  {Julian}}, \ and\ \bibinfo {author} {\bibfnamefont {G.~G.}\ \bibnamefont
  {Lonzarich}},\ }\href@noop {} {\bibfield  {journal} {\bibinfo  {journal}
  {Physical Review B}\ }\textbf {\bibinfo {volume} {55}},\ \bibinfo {pages}
  {8330} (\bibinfo {year} {1997})}\BibitemShut {NoStop}%
\bibitem [{\citenamefont {Schroder}\ \emph {et~al.}(2000)\citenamefont
  {Schroder}, \citenamefont {Aeppli}, \citenamefont {Coldea}, \citenamefont
  {Adams}, \citenamefont {Stockert}, \citenamefont {Lohneysen}, \citenamefont
  {Bucher}, \citenamefont {Ramazashvili},\ and\ \citenamefont
  {Coleman}}]{Schroder2000}%
  \BibitemOpen
  \bibfield  {author} {\bibinfo {author} {\bibfnamefont {A.}~\bibnamefont
  {Schroder}}, \bibinfo {author} {\bibfnamefont {G.}~\bibnamefont {Aeppli}},
  \bibinfo {author} {\bibfnamefont {R.}~\bibnamefont {Coldea}}, \bibinfo
  {author} {\bibfnamefont {M.}~\bibnamefont {Adams}}, \bibinfo {author}
  {\bibfnamefont {O.}~\bibnamefont {Stockert}}, \bibinfo {author}
  {\bibfnamefont {H.~v.}\ \bibnamefont {Lohneysen}}, \bibinfo {author}
  {\bibfnamefont {E.}~\bibnamefont {Bucher}}, \bibinfo {author} {\bibfnamefont
  {R.}~\bibnamefont {Ramazashvili}}, \ and\ \bibinfo {author} {\bibfnamefont
  {P.}~\bibnamefont {Coleman}},\ }\href@noop {} {\bibfield  {journal} {\bibinfo
   {journal} {Nature}\ }\textbf {\bibinfo {volume} {407}},\ \bibinfo {pages}
  {351} (\bibinfo {year} {2000})}\BibitemShut {NoStop}%
\bibitem [{\citenamefont {Lohneysen}\ \emph {et~al.}(1994)\citenamefont
  {Lohneysen}, \citenamefont {Pietrus}, \citenamefont {Portisch}, \citenamefont
  {Schlager}, \citenamefont {Schroder}, \citenamefont {Sieck},\ and\
  \citenamefont {Trappmann}}]{Lohneysen1994}%
  \BibitemOpen
  \bibfield  {author} {\bibinfo {author} {\bibfnamefont {H.~v.}\ \bibnamefont
  {Lohneysen}}, \bibinfo {author} {\bibfnamefont {T.}~\bibnamefont {Pietrus}},
  \bibinfo {author} {\bibfnamefont {G.}~\bibnamefont {Portisch}}, \bibinfo
  {author} {\bibfnamefont {H.~G.}\ \bibnamefont {Schlager}}, \bibinfo {author}
  {\bibfnamefont {A.}~\bibnamefont {Schroder}}, \bibinfo {author}
  {\bibfnamefont {M.}~\bibnamefont {Sieck}}, \ and\ \bibinfo {author}
  {\bibfnamefont {T.}~\bibnamefont {Trappmann}},\ }\href@noop {} {\bibfield
  {journal} {\bibinfo  {journal} {Physical Review Letters}\ }\textbf {\bibinfo
  {volume} {72}},\ \bibinfo {pages} {3262} (\bibinfo {year}
  {1994})}\BibitemShut {NoStop}%
\bibitem [{\citenamefont {Custers}\ \emph {et~al.}(2003)\citenamefont
  {Custers}, \citenamefont {Gegenwart}, \citenamefont {Wilhelm}, \citenamefont
  {Neumaier}, \citenamefont {Tokiwa}, \citenamefont {Trovarelli}, \citenamefont
  {Geibel}, \citenamefont {Steglich}, \citenamefont {Pepin},\ and\
  \citenamefont {Coleman}}]{Custers2003}%
  \BibitemOpen
  \bibfield  {author} {\bibinfo {author} {\bibfnamefont {J.}~\bibnamefont
  {Custers}}, \bibinfo {author} {\bibfnamefont {P.}~\bibnamefont {Gegenwart}},
  \bibinfo {author} {\bibfnamefont {H.}~\bibnamefont {Wilhelm}}, \bibinfo
  {author} {\bibfnamefont {K.}~\bibnamefont {Neumaier}}, \bibinfo {author}
  {\bibfnamefont {Y.}~\bibnamefont {Tokiwa}}, \bibinfo {author} {\bibfnamefont
  {O.}~\bibnamefont {Trovarelli}}, \bibinfo {author} {\bibfnamefont
  {C.}~\bibnamefont {Geibel}}, \bibinfo {author} {\bibfnamefont
  {F.}~\bibnamefont {Steglich}}, \bibinfo {author} {\bibfnamefont
  {C.}~\bibnamefont {Pepin}}, \ and\ \bibinfo {author} {\bibfnamefont
  {P.}~\bibnamefont {Coleman}},\ }\href@noop {} {\bibfield  {journal} {\bibinfo
   {journal} {Nature}\ }\textbf {\bibinfo {volume} {424}},\ \bibinfo {pages}
  {524} (\bibinfo {year} {2003})}\BibitemShut {NoStop}%
\bibitem [{\citenamefont {Pfleiderer}\ and\ \citenamefont
  {Huxley}(2002)}]{Pfleiderer2002}%
  \BibitemOpen
  \bibfield  {author} {\bibinfo {author} {\bibfnamefont {C.}~\bibnamefont
  {Pfleiderer}}\ and\ \bibinfo {author} {\bibfnamefont {A.~D.}\ \bibnamefont
  {Huxley}},\ }\href@noop {} {\bibfield  {journal} {\bibinfo  {journal}
  {Physical Review Letters}\ }\textbf {\bibinfo {volume} {89}},\ \bibinfo
  {pages} {147005} (\bibinfo {year} {2002})}\BibitemShut {NoStop}%
\bibitem [{\citenamefont {Belitz}\ \emph {et~al.}(1999)\citenamefont {Belitz},
  \citenamefont {Kirkpatrick},\ and\ \citenamefont {Vojta}}]{Belitz1999}%
  \BibitemOpen
  \bibfield  {author} {\bibinfo {author} {\bibfnamefont {D.}~\bibnamefont
  {Belitz}}, \bibinfo {author} {\bibfnamefont {T.~R.}\ \bibnamefont
  {Kirkpatrick}}, \ and\ \bibinfo {author} {\bibfnamefont {T.}~\bibnamefont
  {Vojta}},\ }\href@noop {} {\bibfield  {journal} {\bibinfo  {journal}
  {Physical Review Letters}\ }\textbf {\bibinfo {volume} {82}},\ \bibinfo
  {pages} {4707} (\bibinfo {year} {1999})}\BibitemShut {NoStop}%
\bibitem [{\citenamefont {Mineev}(2011)}]{Mineev2011}%
  \BibitemOpen
  \bibfield  {author} {\bibinfo {author} {\bibfnamefont {V.~P.}\ \bibnamefont
  {Mineev}},\ }\href@noop {} {\bibfield  {journal} {\bibinfo  {journal}
  {Comptes Rendus Physique}\ }\textbf {\bibinfo {volume} {12}},\ \bibinfo
  {pages} {567} (\bibinfo {year} {2011})}\BibitemShut {NoStop}%
\bibitem [{\citenamefont {Janoschek}\ \emph {et~al.}(2013)\citenamefont
  {Janoschek}, \citenamefont {Garst}, \citenamefont {Bauer}, \citenamefont
  {Krautscheid}, \citenamefont {Georgii}, \citenamefont {Boni},\ and\
  \citenamefont {Pfleiderer}}]{Janoschek2013}%
  \BibitemOpen
  \bibfield  {author} {\bibinfo {author} {\bibfnamefont {M.}~\bibnamefont
  {Janoschek}}, \bibinfo {author} {\bibfnamefont {M.}~\bibnamefont {Garst}},
  \bibinfo {author} {\bibfnamefont {A.}~\bibnamefont {Bauer}}, \bibinfo
  {author} {\bibfnamefont {P.}~\bibnamefont {Krautscheid}}, \bibinfo {author}
  {\bibfnamefont {R.}~\bibnamefont {Georgii}}, \bibinfo {author} {\bibfnamefont
  {P.}~\bibnamefont {Boni}}, \ and\ \bibinfo {author} {\bibfnamefont
  {C.}~\bibnamefont {Pfleiderer}},\ }\href@noop {} {\bibfield  {journal}
  {\bibinfo  {journal} {Physical Review B}\ }\textbf {\bibinfo {volume} {87}},\
  \bibinfo {pages} {134407} (\bibinfo {year} {2013})}\BibitemShut {NoStop}%
\bibitem [{\citenamefont {Menovsky}\ \emph {et~al.}(1983)\citenamefont
  {Menovsky}, \citenamefont {de~Boer}, \citenamefont {Frings},\ and\
  \citenamefont {Franse}}]{Menovsky1983}%
  \BibitemOpen
  \bibfield  {author} {\bibinfo {author} {\bibfnamefont {A.}~\bibnamefont
  {Menovsky}}, \bibinfo {author} {\bibfnamefont {F.~R.}\ \bibnamefont
  {de~Boer}}, \bibinfo {author} {\bibfnamefont {P.~H.}\ \bibnamefont {Frings}},
  \ and\ \bibinfo {author} {\bibfnamefont {J.~J.~M.}\ \bibnamefont {Franse}},\
  }in\ \href@noop {} {\emph {\bibinfo {booktitle} {High Field Magnetism}}},\
  \bibinfo {editor} {edited by\ \bibinfo {editor} {\bibfnamefont
  {M.}~\bibnamefont {Date}}}\ (\bibinfo  {publisher} {Elsevier},\ \bibinfo
  {address} {Amsterdam},\ \bibinfo {year} {1983})\ pp.\ \bibinfo {pages}
  {189--191}\BibitemShut {NoStop}%
\bibitem [{\citenamefont {Sutherland}\ \emph {et~al.}(2012)\citenamefont
  {Sutherland}, \citenamefont {Smith}, \citenamefont {Marcano}, \citenamefont
  {Zou}, \citenamefont {Rowley}, \citenamefont {Grosche}, \citenamefont
  {Kimura}, \citenamefont {Hayden}, \citenamefont {Takashima}, \citenamefont
  {Nohara},\ and\ \citenamefont {Takagi}}]{Sutherland2012}%
  \BibitemOpen
  \bibfield  {author} {\bibinfo {author} {\bibfnamefont {M.}~\bibnamefont
  {Sutherland}}, \bibinfo {author} {\bibfnamefont {R.~P.}\ \bibnamefont
  {Smith}}, \bibinfo {author} {\bibfnamefont {N.}~\bibnamefont {Marcano}},
  \bibinfo {author} {\bibfnamefont {Y.}~\bibnamefont {Zou}}, \bibinfo {author}
  {\bibfnamefont {S.~E.}\ \bibnamefont {Rowley}}, \bibinfo {author}
  {\bibfnamefont {F.~M.}\ \bibnamefont {Grosche}}, \bibinfo {author}
  {\bibfnamefont {N.}~\bibnamefont {Kimura}}, \bibinfo {author} {\bibfnamefont
  {S.~M.}\ \bibnamefont {Hayden}}, \bibinfo {author} {\bibfnamefont
  {S.}~\bibnamefont {Takashima}}, \bibinfo {author} {\bibfnamefont
  {M.}~\bibnamefont {Nohara}}, \ and\ \bibinfo {author} {\bibfnamefont
  {H.}~\bibnamefont {Takagi}},\ }\href@noop {} {\bibfield  {journal} {\bibinfo
  {journal} {Physical Review B}\ }\textbf {\bibinfo {volume} {85}},\ \bibinfo
  {pages} {035118} (\bibinfo {year} {2012})}\BibitemShut {NoStop}%
\bibitem [{\citenamefont {Aoki}\ and\ \citenamefont
  {Flouquet}(2012)}]{Aoki2012}%
  \BibitemOpen
  \bibfield  {author} {\bibinfo {author} {\bibfnamefont {D.}~\bibnamefont
  {Aoki}}\ and\ \bibinfo {author} {\bibfnamefont {J.}~\bibnamefont
  {Flouquet}},\ }\href@noop {} {\bibfield  {journal} {\bibinfo  {journal}
  {Journal of the Physical Society of Japan}\ }\textbf {\bibinfo {volume}
  {81}},\ \bibinfo {pages} {011003} (\bibinfo {year} {2012})}\BibitemShut
  {NoStop}%
\bibitem [{\citenamefont {Gasparini}\ \emph
  {et~al.}(2010{\natexlab{a}})\citenamefont {Gasparini}, \citenamefont {Huang},
  \citenamefont {Huy}, \citenamefont {Klaasse}, \citenamefont {Naka},
  \citenamefont {Slooten},\ and\ \citenamefont {de~Visser}}]{Gasparini2010a}%
  \BibitemOpen
  \bibfield  {author} {\bibinfo {author} {\bibfnamefont {A.}~\bibnamefont
  {Gasparini}}, \bibinfo {author} {\bibfnamefont {Y.~K.}\ \bibnamefont
  {Huang}}, \bibinfo {author} {\bibfnamefont {N.~T.}\ \bibnamefont {Huy}},
  \bibinfo {author} {\bibfnamefont {J.~C.~P.}\ \bibnamefont {Klaasse}},
  \bibinfo {author} {\bibfnamefont {T.}~\bibnamefont {Naka}}, \bibinfo {author}
  {\bibfnamefont {E.}~\bibnamefont {Slooten}}, \ and\ \bibinfo {author}
  {\bibfnamefont {A.}~\bibnamefont {de~Visser}},\ }\href@noop {} {\bibfield
  {journal} {\bibinfo  {journal} {Journal of Low Temperature Physics}\ }\textbf
  {\bibinfo {volume} {161}},\ \bibinfo {pages} {134} (\bibinfo {year}
  {2010}{\natexlab{a}})}\BibitemShut {NoStop}%
\bibitem [{\citenamefont {Aoki}\ \emph {et~al.}(2009)\citenamefont {Aoki},
  \citenamefont {Matsuda}, \citenamefont {Taufour}, \citenamefont {Hassinger},
  \citenamefont {Knebel},\ and\ \citenamefont {Flouquet}}]{Aoki2009}%
  \BibitemOpen
  \bibfield  {author} {\bibinfo {author} {\bibfnamefont {D.}~\bibnamefont
  {Aoki}}, \bibinfo {author} {\bibfnamefont {T.~D.}\ \bibnamefont {Matsuda}},
  \bibinfo {author} {\bibfnamefont {V.}~\bibnamefont {Taufour}}, \bibinfo
  {author} {\bibfnamefont {E.}~\bibnamefont {Hassinger}}, \bibinfo {author}
  {\bibfnamefont {G.}~\bibnamefont {Knebel}}, \ and\ \bibinfo {author}
  {\bibfnamefont {J.}~\bibnamefont {Flouquet}},\ }\href@noop {} {\bibfield
  {journal} {\bibinfo  {journal} {Journal of the Physical Society of Japan}\
  }\textbf {\bibinfo {volume} {78}},\ \bibinfo {pages} {113709} (\bibinfo
  {year} {2009})}\BibitemShut {NoStop}%
\bibitem [{\citenamefont {Pospisil}\ \emph {et~al.}(2009)\citenamefont
  {Pospisil}, \citenamefont {Vejpravova}, \citenamefont {Divis},\ and\
  \citenamefont {Sechovsky}}]{Pospisil2009}%
  \BibitemOpen
  \bibfield  {author} {\bibinfo {author} {\bibfnamefont {J.}~\bibnamefont
  {Pospisil}}, \bibinfo {author} {\bibfnamefont {J.~P.}\ \bibnamefont
  {Vejpravova}}, \bibinfo {author} {\bibfnamefont {M.}~\bibnamefont {Divis}}, \
  and\ \bibinfo {author} {\bibfnamefont {V.}~\bibnamefont {Sechovsky}},\
  }\href@noop {} {\bibfield  {journal} {\bibinfo  {journal} {Journal of Applied
  Physics}\ }\textbf {\bibinfo {volume} {105}},\ \bibinfo {pages} {07E114}
  (\bibinfo {year} {2009})}\BibitemShut {NoStop}%
\bibitem [{\citenamefont {Troc}\ and\ \citenamefont {Tran}(1988)}]{Troc1988}%
  \BibitemOpen
  \bibfield  {author} {\bibinfo {author} {\bibfnamefont {R.}~\bibnamefont
  {Troc}}\ and\ \bibinfo {author} {\bibfnamefont {V.~H.}\ \bibnamefont
  {Tran}},\ }\href@noop {} {\bibfield  {journal} {\bibinfo  {journal} {Journal
  of Magnetism and Magnetic Materials}\ }\textbf {\bibinfo {volume} {73}},\
  \bibinfo {pages} {389} (\bibinfo {year} {1988})}\BibitemShut {NoStop}%
\bibitem [{\citenamefont {Huy}\ and\ \citenamefont
  {de~Visser}(2009)}]{Huy2009}%
  \BibitemOpen
  \bibfield  {author} {\bibinfo {author} {\bibfnamefont {N.~T.}\ \bibnamefont
  {Huy}}\ and\ \bibinfo {author} {\bibfnamefont {A.}~\bibnamefont
  {de~Visser}},\ }\href@noop {} {\bibfield  {journal} {\bibinfo  {journal}
  {Solid State Communications}\ }\textbf {\bibinfo {volume} {149}},\ \bibinfo
  {pages} {703} (\bibinfo {year} {2009})}\BibitemShut {NoStop}%
\bibitem [{\citenamefont {Huy}\ \emph {et~al.}(2007{\natexlab{b}})\citenamefont
  {Huy}, \citenamefont {Gasparini}, \citenamefont {Klaasse}, \citenamefont
  {de~Visser}, \citenamefont {Sakarya},\ and\ \citenamefont {van
  Dijk}}]{Huy2007b}%
  \BibitemOpen
  \bibfield  {author} {\bibinfo {author} {\bibfnamefont {N.~T.}\ \bibnamefont
  {Huy}}, \bibinfo {author} {\bibfnamefont {A.}~\bibnamefont {Gasparini}},
  \bibinfo {author} {\bibfnamefont {J.~C.~P.}\ \bibnamefont {Klaasse}},
  \bibinfo {author} {\bibfnamefont {A.}~\bibnamefont {de~Visser}}, \bibinfo
  {author} {\bibfnamefont {S.}~\bibnamefont {Sakarya}}, \ and\ \bibinfo
  {author} {\bibfnamefont {N.~H.}\ \bibnamefont {van Dijk}},\ }\href@noop {}
  {\bibfield  {journal} {\bibinfo  {journal} {Physical Review B}\ }\textbf
  {\bibinfo {volume} {75}},\ \bibinfo {pages} {212405} (\bibinfo {year}
  {2007}{\natexlab{b}})}\BibitemShut {NoStop}%
\bibitem [{\citenamefont {Pospisil}\ \emph {et~al.}(2011)\citenamefont
  {Pospisil}, \citenamefont {Prokes}, \citenamefont {Reehuis}, \citenamefont
  {Tovar}, \citenamefont {Poltierova~Vejpravova}, \citenamefont {Prokleska},\
  and\ \citenamefont {Sechovsky}}]{Pospisil2011}%
  \BibitemOpen
  \bibfield  {author} {\bibinfo {author} {\bibfnamefont {J.}~\bibnamefont
  {Pospisil}}, \bibinfo {author} {\bibfnamefont {K.}~\bibnamefont {Prokes}},
  \bibinfo {author} {\bibfnamefont {M.}~\bibnamefont {Reehuis}}, \bibinfo
  {author} {\bibfnamefont {M.}~\bibnamefont {Tovar}}, \bibinfo {author}
  {\bibfnamefont {J.}~\bibnamefont {Poltierova~Vejpravova}}, \bibinfo {author}
  {\bibfnamefont {J.}~\bibnamefont {Prokleska}}, \ and\ \bibinfo {author}
  {\bibfnamefont {V.}~\bibnamefont {Sechovsky}},\ }\href@noop {} {\bibfield
  {journal} {\bibinfo  {journal} {Journal of the Physical Society of Japan}\
  }\textbf {\bibinfo {volume} {80}},\ \bibinfo {pages} {084709} (\bibinfo
  {year} {2011})}\BibitemShut {NoStop}%
\bibitem [{\citenamefont {Rietveld}(1969)}]{Rietveld1969}%
  \BibitemOpen
  \bibfield  {author} {\bibinfo {author} {\bibfnamefont {H.}~\bibnamefont
  {Rietveld}},\ }\href@noop {} {\bibfield  {journal} {\bibinfo  {journal}
  {Journal of Applied Crystallography}\ }\textbf {\bibinfo {volume} {2}},\
  \bibinfo {pages} {65} (\bibinfo {year} {1969})}\BibitemShut {NoStop}%
\bibitem [{\citenamefont {Rodriguez-Carvajal}(1993)}]{Rodriguez-Carvajal1993}%
  \BibitemOpen
  \bibfield  {author} {\bibinfo {author} {\bibfnamefont {J.}~\bibnamefont
  {Rodriguez-Carvajal}},\ }\href@noop {} {\bibfield  {journal} {\bibinfo
  {journal} {Physica B: Condensed Matter}\ }\textbf {\bibinfo {volume} {192}},\
  \bibinfo {pages} {55} (\bibinfo {year} {1993})}\BibitemShut {NoStop}%
\bibitem [{\citenamefont {Roisnel}\ and\ \citenamefont
  {Rodriguez-Carvajal}()}]{Roisnel2000}%
  \BibitemOpen
  \bibfield  {author} {\bibinfo {author} {\bibfnamefont {T.}~\bibnamefont
  {Roisnel}}\ and\ \bibinfo {author} {\bibfnamefont {J.}~\bibnamefont
  {Rodriguez-Carvajal}}\ }(\bibinfo  {publisher} {Trans Tech
  Publications})\BibitemShut {NoStop}%
\bibitem [{\citenamefont {Canepa}\ \emph {et~al.}(1996)\citenamefont {Canepa},
  \citenamefont {Manfrinetti}, \citenamefont {Pani},\ and\ \citenamefont
  {Palenzona}}]{Canepa1996}%
  \BibitemOpen
  \bibfield  {author} {\bibinfo {author} {\bibfnamefont {F.}~\bibnamefont
  {Canepa}}, \bibinfo {author} {\bibfnamefont {P.}~\bibnamefont {Manfrinetti}},
  \bibinfo {author} {\bibfnamefont {M.}~\bibnamefont {Pani}}, \ and\ \bibinfo
  {author} {\bibfnamefont {A.}~\bibnamefont {Palenzona}},\ }\href@noop {}
  {\bibfield  {journal} {\bibinfo  {journal} {Journal of Alloys and Compounds}\
  }\textbf {\bibinfo {volume} {234}},\ \bibinfo {pages} {225} (\bibinfo {year}
  {1996})}\BibitemShut {NoStop}%
\bibitem [{\citenamefont {Perdew}\ and\ \citenamefont
  {Wang}(1992)}]{Perdew1992}%
  \BibitemOpen
  \bibfield  {author} {\bibinfo {author} {\bibfnamefont {J.~P.}\ \bibnamefont
  {Perdew}}\ and\ \bibinfo {author} {\bibfnamefont {Y.}~\bibnamefont {Wang}},\
  }\href@noop {} {\bibfield  {journal} {\bibinfo  {journal} {Physical Review
  B}\ }\textbf {\bibinfo {volume} {45}},\ \bibinfo {pages} {13244} (\bibinfo
  {year} {1992})}\BibitemShut {NoStop}%
\bibitem [{\citenamefont {Perdew}\ \emph {et~al.}(1996)\citenamefont {Perdew},
  \citenamefont {Burke},\ and\ \citenamefont {Ernzerhof}}]{Perdew1996}%
  \BibitemOpen
  \bibfield  {author} {\bibinfo {author} {\bibfnamefont {J.~P.}\ \bibnamefont
  {Perdew}}, \bibinfo {author} {\bibfnamefont {K.}~\bibnamefont {Burke}}, \
  and\ \bibinfo {author} {\bibfnamefont {M.}~\bibnamefont {Ernzerhof}},\
  }\href@noop {} {\bibfield  {journal} {\bibinfo  {journal} {Physical Review
  Letters}\ }\textbf {\bibinfo {volume} {77}},\ \bibinfo {pages} {3865}
  (\bibinfo {year} {1996})}\BibitemShut {NoStop}%
\bibitem [{\citenamefont {Schwarz}\ \emph {et~al.}(2002)\citenamefont
  {Schwarz}, \citenamefont {Blaha},\ and\ \citenamefont
  {Madsen}}]{schwarz2002}%
  \BibitemOpen
  \bibfield  {author} {\bibinfo {author} {\bibfnamefont {K.}~\bibnamefont
  {Schwarz}}, \bibinfo {author} {\bibfnamefont {P.}~\bibnamefont {Blaha}}, \
  and\ \bibinfo {author} {\bibfnamefont {G.~K.~H.}\ \bibnamefont {Madsen}},\
  }\href@noop {} {\bibfield  {journal} {\bibinfo  {journal} {Computer Physics
  Communications}\ }\textbf {\bibinfo {volume} {147}},\ \bibinfo {pages} {71}
  (\bibinfo {year} {2002})}\BibitemShut {NoStop}%
\bibitem [{\citenamefont {Vegard}(1921)}]{Vegard1921}%
  \BibitemOpen
  \bibfield  {author} {\bibinfo {author} {\bibfnamefont {L.}~\bibnamefont
  {Vegard}},\ }\href@noop {} {\bibfield  {journal} {\bibinfo  {journal}
  {Zeitschrift fur Physik A Hadrons and Nuclei}\ }\textbf {\bibinfo {volume}
  {5}},\ \bibinfo {pages} {17} (\bibinfo {year} {1921})}\BibitemShut {NoStop}%
\bibitem [{\citenamefont {Cordero}\ \emph {et~al.}(2008)\citenamefont
  {Cordero}, \citenamefont {Gomez}, \citenamefont {Platero-Prats},
  \citenamefont {Reves}, \citenamefont {Echeverria}, \citenamefont {Cremades},
  \citenamefont {Barragan},\ and\ \citenamefont {Alvarez}}]{Cordero2008}%
  \BibitemOpen
  \bibfield  {author} {\bibinfo {author} {\bibfnamefont {B.}~\bibnamefont
  {Cordero}}, \bibinfo {author} {\bibfnamefont {V.}~\bibnamefont {Gomez}},
  \bibinfo {author} {\bibfnamefont {A.~E.}\ \bibnamefont {Platero-Prats}},
  \bibinfo {author} {\bibfnamefont {M.}~\bibnamefont {Reves}}, \bibinfo
  {author} {\bibfnamefont {J.}~\bibnamefont {Echeverria}}, \bibinfo {author}
  {\bibfnamefont {E.}~\bibnamefont {Cremades}}, \bibinfo {author}
  {\bibfnamefont {F.}~\bibnamefont {Barragan}}, \ and\ \bibinfo {author}
  {\bibfnamefont {S.}~\bibnamefont {Alvarez}},\ }\href@noop {} {\bibfield
  {journal} {\bibinfo  {journal} {Dalton Transactions}\ }\textbf {\bibinfo
  {volume} {0}},\ \bibinfo {pages} {2832} (\bibinfo {year} {2008})}\BibitemShut
  {NoStop}%
\bibitem [{\citenamefont {Arrott}(1957)}]{Arrott1957}%
  \BibitemOpen
  \bibfield  {author} {\bibinfo {author} {\bibfnamefont {A.}~\bibnamefont
  {Arrott}},\ }\href@noop {} {\bibfield  {journal} {\bibinfo  {journal}
  {Physical Review}\ }\textbf {\bibinfo {volume} {108}},\ \bibinfo {pages}
  {1394} (\bibinfo {year} {1957})}\BibitemShut {NoStop}%
\bibitem [{\citenamefont {Valiska}\ \emph {et~al.}(2015)\citenamefont
  {Valiska}, \citenamefont {Pospisil}, \citenamefont {Stunault}, \citenamefont
  {Takeda}, \citenamefont {Gillon}, \citenamefont {Haga}, \citenamefont
  {Prokes}, \citenamefont {Abd-Elmeguid}, \citenamefont {Nenert}, \citenamefont
  {Okane}, \citenamefont {Yamagami}, \citenamefont {Chapon}, \citenamefont
  {Gukasov}, \citenamefont {Cousson}, \citenamefont {Yamamoto},\ and\
  \citenamefont {Sechovsky}}]{Valiska2015}%
  \BibitemOpen
  \bibfield  {author} {\bibinfo {author} {\bibfnamefont {M.}~\bibnamefont
  {Valiska}}, \bibinfo {author} {\bibfnamefont {J.}~\bibnamefont {Pospisil}},
  \bibinfo {author} {\bibfnamefont {A.}~\bibnamefont {Stunault}}, \bibinfo
  {author} {\bibfnamefont {Y.}~\bibnamefont {Takeda}}, \bibinfo {author}
  {\bibfnamefont {B.}~\bibnamefont {Gillon}}, \bibinfo {author} {\bibfnamefont
  {Y.}~\bibnamefont {Haga}}, \bibinfo {author} {\bibfnamefont {K.}~\bibnamefont
  {Prokes}}, \bibinfo {author} {\bibfnamefont {M.~M.}\ \bibnamefont
  {Abd-Elmeguid}}, \bibinfo {author} {\bibfnamefont {G.}~\bibnamefont
  {Nenert}}, \bibinfo {author} {\bibfnamefont {T.}~\bibnamefont {Okane}},
  \bibinfo {author} {\bibfnamefont {H.}~\bibnamefont {Yamagami}}, \bibinfo
  {author} {\bibfnamefont {L.}~\bibnamefont {Chapon}}, \bibinfo {author}
  {\bibfnamefont {A.}~\bibnamefont {Gukasov}}, \bibinfo {author} {\bibfnamefont
  {A.}~\bibnamefont {Cousson}}, \bibinfo {author} {\bibfnamefont
  {E.}~\bibnamefont {Yamamoto}}, \ and\ \bibinfo {author} {\bibfnamefont
  {V.}~\bibnamefont {Sechovsky}},\ }\href@noop {} {} (\bibinfo {year} {2015}),\
  \bibinfo {note} {arXiv:1504.05645}\BibitemShut {NoStop}%
\bibitem [{\citenamefont {Huang}\ \emph {et~al.}(2013)\citenamefont {Huang},
  \citenamefont {Hamlin}, \citenamefont {Baumbach}, \citenamefont {Janoschek},
  \citenamefont {Kanchanavatee}, \citenamefont {Zocco}, \citenamefont
  {Ronning},\ and\ \citenamefont {Maple}}]{Huang2013}%
  \BibitemOpen
  \bibfield  {author} {\bibinfo {author} {\bibfnamefont {K.}~\bibnamefont
  {Huang}}, \bibinfo {author} {\bibfnamefont {J.~J.}\ \bibnamefont {Hamlin}},
  \bibinfo {author} {\bibfnamefont {R.~E.}\ \bibnamefont {Baumbach}}, \bibinfo
  {author} {\bibfnamefont {M.}~\bibnamefont {Janoschek}}, \bibinfo {author}
  {\bibfnamefont {N.}~\bibnamefont {Kanchanavatee}}, \bibinfo {author}
  {\bibfnamefont {D.~A.}\ \bibnamefont {Zocco}}, \bibinfo {author}
  {\bibfnamefont {F.}~\bibnamefont {Ronning}}, \ and\ \bibinfo {author}
  {\bibfnamefont {M.~B.}\ \bibnamefont {Maple}},\ }\href@noop {} {\bibfield
  {journal} {\bibinfo  {journal} {Physical Review B}\ }\textbf {\bibinfo
  {volume} {87}},\ \bibinfo {pages} {054513} (\bibinfo {year}
  {2013})}\BibitemShut {NoStop}%
\bibitem [{\citenamefont {Prokleska}\ \emph {et~al.}(2010)\citenamefont
  {Prokleska}, \citenamefont {Pospisil}, \citenamefont {Vejpravova~Poltierova},
  \citenamefont {Sechovsky},\ and\ \citenamefont {Sebek}}]{Prokleska2010}%
  \BibitemOpen
  \bibfield  {author} {\bibinfo {author} {\bibfnamefont {J.}~\bibnamefont
  {Prokleska}}, \bibinfo {author} {\bibfnamefont {J.}~\bibnamefont {Pospisil}},
  \bibinfo {author} {\bibfnamefont {J.}~\bibnamefont {Vejpravova~Poltierova}},
  \bibinfo {author} {\bibfnamefont {V.}~\bibnamefont {Sechovsky}}, \ and\
  \bibinfo {author} {\bibfnamefont {J.}~\bibnamefont {Sebek}},\ }\href@noop {}
  {\bibfield  {journal} {\bibinfo  {journal} {Journal of Physics: Conference
  Series}\ }\textbf {\bibinfo {volume} {200}},\ \bibinfo {pages} {012161}
  (\bibinfo {year} {2010})}\BibitemShut {NoStop}%
\bibitem [{\citenamefont {Millis}(1993)}]{Millis1993}%
  \BibitemOpen
  \bibfield  {author} {\bibinfo {author} {\bibfnamefont {A.~J.}\ \bibnamefont
  {Millis}},\ }\href@noop {} {\bibfield  {journal} {\bibinfo  {journal}
  {Physical Review B}\ }\textbf {\bibinfo {volume} {48}},\ \bibinfo {pages}
  {7183} (\bibinfo {year} {1993})}\BibitemShut {NoStop}%
\bibitem [{\citenamefont {Hertz}(1976)}]{Hertz1976}%
  \BibitemOpen
  \bibfield  {author} {\bibinfo {author} {\bibfnamefont {J.~A.}\ \bibnamefont
  {Hertz}},\ }\href@noop {} {\bibfield  {journal} {\bibinfo  {journal}
  {Physical Review B}\ }\textbf {\bibinfo {volume} {14}},\ \bibinfo {pages}
  {1165} (\bibinfo {year} {1976})}\BibitemShut {NoStop}%
\bibitem [{\citenamefont {Stewart}(2001)}]{Stewart2001}%
  \BibitemOpen
  \bibfield  {author} {\bibinfo {author} {\bibfnamefont {G.~R.}\ \bibnamefont
  {Stewart}},\ }\href@noop {} {\bibfield  {journal} {\bibinfo  {journal}
  {Reviews of Modern Physics}\ }\textbf {\bibinfo {volume} {73}},\ \bibinfo
  {pages} {797} (\bibinfo {year} {2001})}\BibitemShut {NoStop}%
\bibitem [{\citenamefont {Gasparini}\ \emph
  {et~al.}(2010{\natexlab{b}})\citenamefont {Gasparini}, \citenamefont {Huang},
  \citenamefont {Hartbaum}, \citenamefont {von Lohneysen},\ and\ \citenamefont
  {de~Visser}}]{Gasparini2010}%
  \BibitemOpen
  \bibfield  {author} {\bibinfo {author} {\bibfnamefont {A.}~\bibnamefont
  {Gasparini}}, \bibinfo {author} {\bibfnamefont {Y.~K.}\ \bibnamefont
  {Huang}}, \bibinfo {author} {\bibfnamefont {J.}~\bibnamefont {Hartbaum}},
  \bibinfo {author} {\bibfnamefont {H.}~\bibnamefont {von Lohneysen}}, \ and\
  \bibinfo {author} {\bibfnamefont {A.}~\bibnamefont {de~Visser}},\ }\href@noop
  {} {\bibfield  {journal} {\bibinfo  {journal} {Physical Review B}\ }\textbf
  {\bibinfo {volume} {82}},\ \bibinfo {pages} {052502} (\bibinfo {year}
  {2010}{\natexlab{b}})}\BibitemShut {NoStop}%
\bibitem [{\citenamefont {Divis}(2008)}]{Divis2008}%
  \BibitemOpen
  \bibfield  {author} {\bibinfo {author} {\bibfnamefont {M.}~\bibnamefont
  {Divis}},\ }\href@noop {} {\bibfield  {journal} {\bibinfo  {journal} {Physica
  B-Condensed Matter}\ }\textbf {\bibinfo {volume} {403}},\ \bibinfo {pages}
  {2505} (\bibinfo {year} {2008})}\BibitemShut {NoStop}%
\bibitem [{\citenamefont {Neto}\ \emph {et~al.}(2013)\citenamefont {Neto},
  \citenamefont {Neto}, \citenamefont {Kim},\ and\ \citenamefont
  {Stewart}}]{Silva2013}%
  \BibitemOpen
  \bibfield  {author} {\bibinfo {author} {\bibfnamefont {M.~B.~S.}\
  \bibnamefont {Neto}}, \bibinfo {author} {\bibfnamefont {A.~H.~C.}\
  \bibnamefont {Neto}}, \bibinfo {author} {\bibfnamefont {J.~S.}\ \bibnamefont
  {Kim}}, \ and\ \bibinfo {author} {\bibfnamefont {G.~R.}\ \bibnamefont
  {Stewart}},\ }\href@noop {} {\bibfield  {journal} {\bibinfo  {journal}
  {Journal of Physics: Condensed Matter}\ }\textbf {\bibinfo {volume} {25}},\
  \bibinfo {pages} {025601} (\bibinfo {year} {2013})}\BibitemShut {NoStop}%
\bibitem [{\citenamefont {Batista}\ \emph {et~al.}(2002)\citenamefont
  {Batista}, \citenamefont {Bonca},\ and\ \citenamefont
  {Gubernatis}}]{Batista2002}%
  \BibitemOpen
  \bibfield  {author} {\bibinfo {author} {\bibfnamefont {C.~D.}\ \bibnamefont
  {Batista}}, \bibinfo {author} {\bibfnamefont {J.}~\bibnamefont {Bonca}}, \
  and\ \bibinfo {author} {\bibfnamefont {J.~E.}\ \bibnamefont {Gubernatis}},\
  }\href@noop {} {\bibfield  {journal} {\bibinfo  {journal} {Physical Review
  Letters}\ }\textbf {\bibinfo {volume} {88}},\ \bibinfo {pages} {187203}
  (\bibinfo {year} {2002})}\BibitemShut {NoStop}%
\bibitem [{\citenamefont {Batista}\ \emph {et~al.}(2003)\citenamefont
  {Batista}, \citenamefont {Bonca},\ and\ \citenamefont
  {Gubernatis}}]{Batista2003}%
  \BibitemOpen
  \bibfield  {author} {\bibinfo {author} {\bibfnamefont {C.~D.}\ \bibnamefont
  {Batista}}, \bibinfo {author} {\bibfnamefont {J.}~\bibnamefont {Bonca}}, \
  and\ \bibinfo {author} {\bibfnamefont {J.~E.}\ \bibnamefont {Gubernatis}},\
  }\href@noop {} {\bibfield  {journal} {\bibinfo  {journal} {Physical Review
  B}\ }\textbf {\bibinfo {volume} {68}},\ \bibinfo {pages} {214430} (\bibinfo
  {year} {2003})}\BibitemShut {NoStop}%
\bibitem [{\citenamefont {Gasche}\ \emph {et~al.}(1995)\citenamefont {Gasche},
  \citenamefont {Brooks},\ and\ \citenamefont {Johansson}}]{Gasche1995}%
  \BibitemOpen
  \bibfield  {author} {\bibinfo {author} {\bibfnamefont {T.}~\bibnamefont
  {Gasche}}, \bibinfo {author} {\bibfnamefont {M.~S.~S.}\ \bibnamefont
  {Brooks}}, \ and\ \bibinfo {author} {\bibfnamefont {B.}~\bibnamefont
  {Johansson}},\ }\href@noop {} {\bibfield  {journal} {\bibinfo  {journal}
  {Journal of Physics: Condensed Matter}\ }\textbf {\bibinfo {volume} {7}},\
  \bibinfo {pages} {9499} (\bibinfo {year} {1995})}\BibitemShut {NoStop}%
\bibitem [{\citenamefont {Newns}\ and\ \citenamefont {Read}(1987)}]{Newns1987}%
  \BibitemOpen
  \bibfield  {author} {\bibinfo {author} {\bibfnamefont {D.~M.}\ \bibnamefont
  {Newns}}\ and\ \bibinfo {author} {\bibfnamefont {N.}~\bibnamefont {Read}},\
  }\href@noop {} {\bibfield  {journal} {\bibinfo  {journal} {Advances in
  Physics}\ }\textbf {\bibinfo {volume} {36}},\ \bibinfo {pages} {799}
  (\bibinfo {year} {1987})}\BibitemShut {NoStop}%
\bibitem [{\citenamefont {Moriya}(2012)}]{Moriya2012}%
  \BibitemOpen
  \bibfield  {author} {\bibinfo {author} {\bibfnamefont {T.}~\bibnamefont
  {Moriya}},\ }\href@noop {} {}\ (\bibinfo  {publisher} {Springer Berlin
  Heidelberg},\ \bibinfo {year} {2012})\BibitemShut {NoStop}%
\bibitem [{\citenamefont {Smith}\ and\ \citenamefont
  {Kmetko}(1983)}]{Smith1983}%
  \BibitemOpen
  \bibfield  {author} {\bibinfo {author} {\bibfnamefont {J.~L.}\ \bibnamefont
  {Smith}}\ and\ \bibinfo {author} {\bibfnamefont {E.~A.}\ \bibnamefont
  {Kmetko}},\ }\href@noop {} {\bibfield  {journal} {\bibinfo  {journal}
  {Journal of the Less Common Metals}\ }\textbf {\bibinfo {volume} {90}},\
  \bibinfo {pages} {83} (\bibinfo {year} {1983})}\BibitemShut {NoStop}%
\bibitem [{\citenamefont {Cooper}\ \emph {et~al.}(1992)\citenamefont {Cooper},
  \citenamefont {Sheng}, \citenamefont {Lim}, \citenamefont {Sanchez-Castro},
  \citenamefont {Kioussis},\ and\ \citenamefont {Wills}}]{Cooper1992}%
  \BibitemOpen
  \bibfield  {author} {\bibinfo {author} {\bibfnamefont {B.~R.}\ \bibnamefont
  {Cooper}}, \bibinfo {author} {\bibfnamefont {Q.~G.}\ \bibnamefont {Sheng}},
  \bibinfo {author} {\bibfnamefont {S.~P.}\ \bibnamefont {Lim}}, \bibinfo
  {author} {\bibfnamefont {C.}~\bibnamefont {Sanchez-Castro}}, \bibinfo
  {author} {\bibfnamefont {N.}~\bibnamefont {Kioussis}}, \ and\ \bibinfo
  {author} {\bibfnamefont {J.~M.}\ \bibnamefont {Wills}},\ }\href@noop {}
  {\bibfield  {journal} {\bibinfo  {journal} {Journal of Magnetism and Magnetic
  Materials}\ }\textbf {\bibinfo {volume} {108}},\ \bibinfo {pages} {10}
  (\bibinfo {year} {1992})}\BibitemShut {NoStop}%
\bibitem [{\citenamefont {Brooks}\ and\ \citenamefont
  {Kelly}(1983)}]{Brooks1983}%
  \BibitemOpen
  \bibfield  {author} {\bibinfo {author} {\bibfnamefont {M.~S.~S.}\
  \bibnamefont {Brooks}}\ and\ \bibinfo {author} {\bibfnamefont {P.~J.}\
  \bibnamefont {Kelly}},\ }\href@noop {} {\bibfield  {journal} {\bibinfo
  {journal} {Physical Review Letters}\ }\textbf {\bibinfo {volume} {51}},\
  \bibinfo {pages} {1708} (\bibinfo {year} {1983})}\BibitemShut {NoStop}%
\bibitem [{\citenamefont {Lander}\ \emph {et~al.}(1991)\citenamefont {Lander},
  \citenamefont {Brooks},\ and\ \citenamefont {Johansson}}]{Lander1991}%
  \BibitemOpen
  \bibfield  {author} {\bibinfo {author} {\bibfnamefont {G.~H.}\ \bibnamefont
  {Lander}}, \bibinfo {author} {\bibfnamefont {M.~S.~S.}\ \bibnamefont
  {Brooks}}, \ and\ \bibinfo {author} {\bibfnamefont {B.}~\bibnamefont
  {Johansson}},\ }\href@noop {} {\bibfield  {journal} {\bibinfo  {journal}
  {Physical Review B}\ }\textbf {\bibinfo {volume} {43}},\ \bibinfo {pages}
  {13672} (\bibinfo {year} {1991})}\BibitemShut {NoStop}%
\bibitem [{\citenamefont {Sechovsky}\ and\ \citenamefont
  {Havela}(1998)}]{Sechovsky1998}%
  \BibitemOpen
  \bibfield  {author} {\bibinfo {author} {\bibfnamefont {V.}~\bibnamefont
  {Sechovsky}}\ and\ \bibinfo {author} {\bibfnamefont {L.}~\bibnamefont
  {Havela}},\ }in\ \href@noop {} {\emph {\bibinfo {booktitle} {Handbook of
  Magnetic Materials}}},\ Vol.\ \bibinfo {volume} {Volume 11},\ \bibinfo
  {editor} {edited by\ \bibinfo {editor} {\bibfnamefont {K.~H.~J.}\
  \bibnamefont {Buschow}}}\ (\bibinfo  {publisher} {Elsevier},\ \bibinfo {year}
  {1998})\ pp.\ \bibinfo {pages} {1--289}\BibitemShut {NoStop}%
\bibitem [{\citenamefont {Sanchez-Castro}\ \emph {et~al.}(1995)\citenamefont
  {Sanchez-Castro}, \citenamefont {Cooper},\ and\ \citenamefont
  {Bedell}}]{Sanchez1995}%
  \BibitemOpen
  \bibfield  {author} {\bibinfo {author} {\bibfnamefont {C.}~\bibnamefont
  {Sanchez-Castro}}, \bibinfo {author} {\bibfnamefont {B.~R.}\ \bibnamefont
  {Cooper}}, \ and\ \bibinfo {author} {\bibfnamefont {K.~S.}\ \bibnamefont
  {Bedell}},\ }\href@noop {} {\bibfield  {journal} {\bibinfo  {journal}
  {Physical Review B}\ }\textbf {\bibinfo {volume} {51}},\ \bibinfo {pages}
  {12506} (\bibinfo {year} {1995})}\BibitemShut {NoStop}%
\bibitem [{\citenamefont {Taupin}\ \emph {et~al.}(2015)\citenamefont {Taupin},
  \citenamefont {Brison}, \citenamefont {Aoki}, \citenamefont {Sanchez},
  \citenamefont {Wilhelm},\ and\ \citenamefont {Rogalev}}]{Taupin2015}%
  \BibitemOpen
  \bibfield  {author} {\bibinfo {author} {\bibfnamefont {M.}~\bibnamefont
  {Taupin}}, \bibinfo {author} {\bibfnamefont {J.-P.}\ \bibnamefont {Brison}},
  \bibinfo {author} {\bibfnamefont {D.}~\bibnamefont {Aoki}}, \bibinfo {author}
  {\bibfnamefont {J.-P.}\ \bibnamefont {Sanchez}}, \bibinfo {author}
  {\bibfnamefont {F.}~\bibnamefont {Wilhelm}}, \ and\ \bibinfo {author}
  {\bibfnamefont {A.}~\bibnamefont {Rogalev}},\ }\href@noop {} {\bibfield
  {journal} {\bibinfo  {journal} {ArXiv e-prints}\ } (\bibinfo {year}
  {2015})}\BibitemShut {NoStop}%
\bibitem [{\citenamefont {Hill}()}]{Hill1970}%
  \BibitemOpen
  \bibfield  {author} {\bibinfo {author} {\bibfnamefont {H.~H.}\ \bibnamefont
  {Hill}},\ }pp.\ \bibinfo {pages} {1 -- 19}\BibitemShut {NoStop}%
\bibitem [{\citenamefont {Wulff}\ \emph {et~al.}(1989)\citenamefont {Wulff},
  \citenamefont {Lander}, \citenamefont {Lebech},\ and\ \citenamefont
  {Delapalme}}]{Wulff1989}%
  \BibitemOpen
  \bibfield  {author} {\bibinfo {author} {\bibfnamefont {M.}~\bibnamefont
  {Wulff}}, \bibinfo {author} {\bibfnamefont {G.~H.}\ \bibnamefont {Lander}},
  \bibinfo {author} {\bibfnamefont {B.}~\bibnamefont {Lebech}}, \ and\ \bibinfo
  {author} {\bibfnamefont {A.}~\bibnamefont {Delapalme}},\ }\href@noop {}
  {\bibfield  {journal} {\bibinfo  {journal} {Physical Review B}\ }\textbf
  {\bibinfo {volume} {39}},\ \bibinfo {pages} {4719} (\bibinfo {year}
  {1989})}\BibitemShut {NoStop}%
\bibitem [{\citenamefont {Troc}\ \emph {et~al.}(1993)\citenamefont {Troc},
  \citenamefont {Tran}, \citenamefont {Vagizov},\ and\ \citenamefont
  {Drulis}}]{Troc1993}%
  \BibitemOpen
  \bibfield  {author} {\bibinfo {author} {\bibfnamefont {R.}~\bibnamefont
  {Troc}}, \bibinfo {author} {\bibfnamefont {V.~H.}\ \bibnamefont {Tran}},
  \bibinfo {author} {\bibfnamefont {F.~G.}\ \bibnamefont {Vagizov}}, \ and\
  \bibinfo {author} {\bibfnamefont {H.}~\bibnamefont {Drulis}},\ }\href@noop {}
  {\bibfield  {journal} {\bibinfo  {journal} {Journal of Alloys and Compounds}\
  }\textbf {\bibinfo {volume} {200}},\ \bibinfo {pages} {37} (\bibinfo {year}
  {1993})}\BibitemShut {NoStop}%
\bibitem [{\citenamefont {Veenhuizen}\ \emph {et~al.}(1988)\citenamefont
  {Veenhuizen}, \citenamefont {de~Boer}, \citenamefont {Menovsky},
  \citenamefont {Sechovsky},\ and\ \citenamefont {Havela}}]{Veenhuizen1988}%
  \BibitemOpen
  \bibfield  {author} {\bibinfo {author} {\bibfnamefont {P.~A.}\ \bibnamefont
  {Veenhuizen}}, \bibinfo {author} {\bibfnamefont {F.~R.}\ \bibnamefont
  {de~Boer}}, \bibinfo {author} {\bibfnamefont {A.~A.}\ \bibnamefont
  {Menovsky}}, \bibinfo {author} {\bibfnamefont {V.}~\bibnamefont {Sechovsky}},
  \ and\ \bibinfo {author} {\bibfnamefont {L.}~\bibnamefont {Havela}},\
  }\href@noop {} {\bibfield  {journal} {\bibinfo  {journal} {Journal de
  Physique}\ }\textbf {\bibinfo {volume} {49}},\ \bibinfo {pages} {485}
  (\bibinfo {year} {1988})}\BibitemShut {NoStop}%
\bibitem [{\citenamefont {Eriksson}\ \emph {et~al.}(1989)\citenamefont
  {Eriksson}, \citenamefont {Johansson},\ and\ \citenamefont
  {Brooks}}]{Eriksson1989}%
  \BibitemOpen
  \bibfield  {author} {\bibinfo {author} {\bibfnamefont {O.}~\bibnamefont
  {Eriksson}}, \bibinfo {author} {\bibfnamefont {B.}~\bibnamefont {Johansson}},
  \ and\ \bibinfo {author} {\bibfnamefont {M.~S.~S.}\ \bibnamefont {Brooks}},\
  }\href@noop {} {\bibfield  {journal} {\bibinfo  {journal} {Journal of
  Physics: Condensed Matter}\ }\textbf {\bibinfo {volume} {1}},\ \bibinfo
  {pages} {4005} (\bibinfo {year} {1989})}\BibitemShut {NoStop}%
\bibitem [{\citenamefont {Shimizu}(1982)}]{Shimizu1982}%
  \BibitemOpen
  \bibfield  {author} {\bibinfo {author} {\bibfnamefont {M.}~\bibnamefont
  {Shimizu}},\ }\href@noop {} {\bibfield  {journal} {\bibinfo  {journal} {J.
  Phys. France}\ }\textbf {\bibinfo {volume} {43}},\ \bibinfo {pages} {155}
  (\bibinfo {year} {1982})}\BibitemShut {NoStop}%
\bibitem [{\citenamefont {Mushnikov}\ \emph {et~al.}(1999)\citenamefont
  {Mushnikov}, \citenamefont {Goto}, \citenamefont {Kamishima}, \citenamefont
  {Yamada}, \citenamefont {Andreev}, \citenamefont {Shiokawa}, \citenamefont
  {Iwao},\ and\ \citenamefont {Sechovsky}}]{Mushnikov1999}%
  \BibitemOpen
  \bibfield  {author} {\bibinfo {author} {\bibfnamefont {N.~V.}\ \bibnamefont
  {Mushnikov}}, \bibinfo {author} {\bibfnamefont {T.}~\bibnamefont {Goto}},
  \bibinfo {author} {\bibfnamefont {K.}~\bibnamefont {Kamishima}}, \bibinfo
  {author} {\bibfnamefont {H.}~\bibnamefont {Yamada}}, \bibinfo {author}
  {\bibfnamefont {A.~V.}\ \bibnamefont {Andreev}}, \bibinfo {author}
  {\bibfnamefont {Y.}~\bibnamefont {Shiokawa}}, \bibinfo {author}
  {\bibfnamefont {A.}~\bibnamefont {Iwao}}, \ and\ \bibinfo {author}
  {\bibfnamefont {V.}~\bibnamefont {Sechovsky}},\ }\href@noop {} {\bibfield
  {journal} {\bibinfo  {journal} {Physical Review B}\ }\textbf {\bibinfo
  {volume} {59}},\ \bibinfo {pages} {6877} (\bibinfo {year}
  {1999})}\BibitemShut {NoStop}%
\bibitem [{\citenamefont {Andreev}\ \emph {et~al.}(1997)\citenamefont
  {Andreev}, \citenamefont {Havela}, \citenamefont {Sechovsky}, \citenamefont
  {Bartashevich}, \citenamefont {Sebek}, \citenamefont {Dremov},\ and\
  \citenamefont {Kozlovskaya}}]{Andreev1997a}%
  \BibitemOpen
  \bibfield  {author} {\bibinfo {author} {\bibfnamefont {A.~V.}\ \bibnamefont
  {Andreev}}, \bibinfo {author} {\bibfnamefont {L.}~\bibnamefont {Havela}},
  \bibinfo {author} {\bibfnamefont {V.}~\bibnamefont {Sechovsky}}, \bibinfo
  {author} {\bibfnamefont {M.~I.}\ \bibnamefont {Bartashevich}}, \bibinfo
  {author} {\bibfnamefont {J.}~\bibnamefont {Sebek}}, \bibinfo {author}
  {\bibfnamefont {R.~V.}\ \bibnamefont {Dremov}}, \ and\ \bibinfo {author}
  {\bibfnamefont {I.~K.}\ \bibnamefont {Kozlovskaya}},\ }\href@noop {}
  {\bibfield  {journal} {\bibinfo  {journal} {Philosophical Magazine B-Physics
  of Condensed Matter Statistical Mechanics Electronic Optical and Magnetic
  Properties}\ }\textbf {\bibinfo {volume} {75}},\ \bibinfo {pages} {827}
  (\bibinfo {year} {1997})}\BibitemShut {NoStop}%
\bibitem [{\citenamefont {Sechovsky}\ \emph {et~al.}(1992)\citenamefont
  {Sechovsky}, \citenamefont {Havela}, \citenamefont {de~Boer}, \citenamefont
  {Veenhuizen}, \citenamefont {Sugiyama}, \citenamefont {Kuroda}, \citenamefont
  {Sugiura}, \citenamefont {Ono}, \citenamefont {Date},\ and\ \citenamefont
  {Yamagishi}}]{Sechovsky1992}%
  \BibitemOpen
  \bibfield  {author} {\bibinfo {author} {\bibfnamefont {V.}~\bibnamefont
  {Sechovsky}}, \bibinfo {author} {\bibfnamefont {L.}~\bibnamefont {Havela}},
  \bibinfo {author} {\bibfnamefont {F.~R.}\ \bibnamefont {de~Boer}}, \bibinfo
  {author} {\bibfnamefont {P.~A.}\ \bibnamefont {Veenhuizen}}, \bibinfo
  {author} {\bibfnamefont {K.}~\bibnamefont {Sugiyama}}, \bibinfo {author}
  {\bibfnamefont {T.}~\bibnamefont {Kuroda}}, \bibinfo {author} {\bibfnamefont
  {E.}~\bibnamefont {Sugiura}}, \bibinfo {author} {\bibfnamefont
  {M.}~\bibnamefont {Ono}}, \bibinfo {author} {\bibfnamefont {M.}~\bibnamefont
  {Date}}, \ and\ \bibinfo {author} {\bibfnamefont {A.}~\bibnamefont
  {Yamagishi}},\ }\href@noop {} {\bibfield  {journal} {\bibinfo  {journal}
  {Physica B: Condensed Matter}\ }\textbf {\bibinfo {volume} {177}},\ \bibinfo
  {pages} {164} (\bibinfo {year} {1992})}\BibitemShut {NoStop}%
\bibitem [{\citenamefont {Mushnikov}\ \emph {et~al.}(2002)\citenamefont
  {Mushnikov}, \citenamefont {Goto}, \citenamefont {Andreev}, \citenamefont
  {Sechovsky},\ and\ \citenamefont {Yamada}}]{Mushnikov2002}%
  \BibitemOpen
  \bibfield  {author} {\bibinfo {author} {\bibfnamefont {N.~V.}\ \bibnamefont
  {Mushnikov}}, \bibinfo {author} {\bibfnamefont {T.}~\bibnamefont {Goto}},
  \bibinfo {author} {\bibfnamefont {A.~V.}\ \bibnamefont {Andreev}}, \bibinfo
  {author} {\bibfnamefont {V.}~\bibnamefont {Sechovsky}}, \ and\ \bibinfo
  {author} {\bibfnamefont {H.}~\bibnamefont {Yamada}},\ }\href@noop {}
  {\bibfield  {journal} {\bibinfo  {journal} {Physical Review B}\ }\textbf
  {\bibinfo {volume} {66}},\ \bibinfo {pages} {064433} (\bibinfo {year}
  {2002})}\BibitemShut {NoStop}%
\bibitem [{\citenamefont {Belitz}\ and\ \citenamefont
  {Kirkpatrick}(2012)}]{Belitz2012}%
  \BibitemOpen
  \bibfield  {author} {\bibinfo {author} {\bibfnamefont {D.}~\bibnamefont
  {Belitz}}\ and\ \bibinfo {author} {\bibfnamefont {T.~R.}\ \bibnamefont
  {Kirkpatrick}},\ }\href@noop {} {} (\bibinfo {year} {2012}),\ \bibinfo {note}
  {arXiv:1204.0873}\BibitemShut {NoStop}%
\bibitem [{\citenamefont {Hattori}\ \emph {et~al.}(2010)\citenamefont
  {Hattori}, \citenamefont {Ishida}, \citenamefont {Nakai}, \citenamefont
  {Ohta}, \citenamefont {Deguchi}, \citenamefont {Sato},\ and\ \citenamefont
  {Satoh}}]{Hattori2010}%
  \BibitemOpen
  \bibfield  {author} {\bibinfo {author} {\bibfnamefont {T.}~\bibnamefont
  {Hattori}}, \bibinfo {author} {\bibfnamefont {K.}~\bibnamefont {Ishida}},
  \bibinfo {author} {\bibfnamefont {Y.}~\bibnamefont {Nakai}}, \bibinfo
  {author} {\bibfnamefont {T.}~\bibnamefont {Ohta}}, \bibinfo {author}
  {\bibfnamefont {K.}~\bibnamefont {Deguchi}}, \bibinfo {author} {\bibfnamefont
  {N.~K.}\ \bibnamefont {Sato}}, \ and\ \bibinfo {author} {\bibfnamefont
  {I.}~\bibnamefont {Satoh}},\ }\href@noop {} {\bibfield  {journal} {\bibinfo
  {journal} {Physica C: Superconductivity}\ }\textbf {\bibinfo {volume}
  {470}},\ \bibinfo {pages} {S561} (\bibinfo {year} {2010})}\BibitemShut
  {NoStop}%
\bibitem [{\citenamefont {Miranda}\ \emph {et~al.}(1997)\citenamefont
  {Miranda}, \citenamefont {Dobrosavljevic},\ and\ \citenamefont
  {Kotliar}}]{Miranda1997}%
  \BibitemOpen
  \bibfield  {author} {\bibinfo {author} {\bibfnamefont {E.}~\bibnamefont
  {Miranda}}, \bibinfo {author} {\bibfnamefont {V.}~\bibnamefont
  {Dobrosavljevic}}, \ and\ \bibinfo {author} {\bibfnamefont {G.}~\bibnamefont
  {Kotliar}},\ }\href@noop {} {\bibfield  {journal} {\bibinfo  {journal}
  {Physical Review Letters}\ }\textbf {\bibinfo {volume} {78}},\ \bibinfo
  {pages} {290} (\bibinfo {year} {1997})}\BibitemShut {NoStop}%
\bibitem [{\citenamefont {Rosch}(1999)}]{Rosch1999}%
  \BibitemOpen
  \bibfield  {author} {\bibinfo {author} {\bibfnamefont {A.}~\bibnamefont
  {Rosch}},\ }\href@noop {} {\bibfield  {journal} {\bibinfo  {journal}
  {Physical Review Letters}\ }\textbf {\bibinfo {volume} {82}},\ \bibinfo
  {pages} {4280} (\bibinfo {year} {1999})}\BibitemShut {NoStop}%
\end{thebibliography}%

\end{document}